\documentclass[ejsv2,noshowframe]{imsart}

\RequirePackage{amsthm,amsmath,amsfonts}
\RequirePackage[numbers]{natbib}
\RequirePackage[colorlinks,citecolor=blue,urlcolor=blue]{hyperref}
\RequirePackage{graphicx}
\RequirePackage{subcaption}
\usepackage[table]{xcolor}
\usepackage{algorithm}
\usepackage{algpseudocode}
\usepackage{mathtools}
\makeatletter
\@ifpackageloaded{newtxmath}{}{}
\makeatother
\usepackage{amssymb}

\theoremstyle{plain}
\newtheorem{assumption}{Assumption}

\arxiv{2010.00000}
\startlocaldefs
\theoremstyle{plain}

\theoremstyle{definition}

\theoremstyle{remark}

\endlocaldefs

\begin{document}
\sloppy
\begin{frontmatter}
\title{Tracking Temporal Evolution of Topological Features in Image Data}
\runtitle{Temporal Evolution of Topological Features}

\begin{aug}
\author[A,B]{\fnms{Susan}~\snm{Glenn}\ead[label=e1]{ssglenn@lanl.gov}},
\author[B]{\fnms{Jessi}~\snm{Cisewski-Kehe}\ead[label=e2]{jjkehe@wisc.edu}}
\author[B]{\fnms{Jun}~\snm{Zhu}\ead[label=e3]{junzhu@wisc.edu}}
\and
\author[C]{\fnms{William M.}~\snm{Bement}\ead[label=e4]{wmbement@wisc.edu}}
\address[A]{Los Alamos National Laboratory,
Los Alamos, NM, USA\printead[presep={,\ }]{e1}}

\address[B]{Department of Statistics,
University of Wisconsin, USA\printead[presep={,\ }]{e2,e3}}

\address[C]{Center for Quantitative Cell Imaging, 
Department of Integrative Biology, 
University of Wisconsin, USA \printead[presep={,\ }]{e4}}
\runauthor{Glenn et al.}
\end{aug}

\begin{abstract}

Topological Data Analysis (TDA) can be used to detect and characterize holes in an image, such as zero-dimensional holes (connected components) or one-dimensional holes (loops). However, there is currently no widely accepted statistical framework for modeling spatiotemporal dependence in the evolution of topological features, such as holes, within a time series of images. We propose a hypothesis testing framework to identify statistically significant topological features of images in space and time, simultaneously. This addition of time may induce higher-dimensional topological features which can be used to establish temporal connections between the lower-dimensional features at each point in time. The temporal evolution of these lower-dimensional features is then represented on a zigzag persistence diagram, as a topological summary statistic focused on time dynamics. We demonstrate that the method effectively captures the emergence and progression of topological features in a study of a series of images of a wounded cell as it repairs. The proposed method outperforms a current approach in a simulation study that includes features of the wound healing process. Since, the wounded cell images exhibit nonlinear, dynamic, spatial, and temporal structures during single-cell repair, they provide a good application for this method.
\end{abstract}



\end{frontmatter}

\section{Introduction}

Topological data analysis (TDA) provides a flexible framework for capturing nonlinear structure in image data by characterizing shape through connectivity and holes. A hole in an image is a topological feature characterized by a specific dimension, such as, a connected component (dimension 0), a loop (dimension 1), or a void (dimension 2), etc. The overall hole structure, which reflects holes across different dimensions, can be summarized using topological summary statistics. There have been many successful applications of TDA to understanding complex patterns in images, from neuroscience to climate science, and much more \citep{r20, r28, r29, r30, r1, r31}. Extending these ideas to temporal data typically involves either computing time-varying topological summary statistics or applying TDA to lower-dimensional embeddings that capture the underlying dynamics. Images, in general, are challenging as they are typically multidimensional with non-linear patterns. Spatiotemporal image data add further complexity due to the combination of both spatial dependencies (e.g., nearby pixels in an image are more related than pixels further away), temporal dependencies (e.g., recent images are more related to present images than those in the more distant past), and their interactions. Despite advances in statistical techniques to disentangle these multi-scale dependencies in non-linear, high-dimensional spaces \cite{r32}, it remains a challenge to detect and characterize topological features of images evolving over time.

Techniques in TDA for tracking temporal dynamics in data include sliding window embeddings, zigzag persistence, crocker plots, multiparameter persistence, and vineyards. However, most were developed for data such as one-dimensional times series, networks, point clouds, or binary images, and do not directly generalize to sequences of grayscale images \citep{r4, r6, r7, r8, r9, r11, r12, r27}. A sliding window embedding maps overlapping time domain windows of a time series to a point-cloud representation and is most commonly used with scalar time series. While some work has extended these techniques to time series of higher-dimensional data structures, such as images or networks, they usually either (i) reduce the images or graphs to a scalar-valued function and then track temporal changes, or (ii) capture only global temporal dynamics throughout a video as opposed to specific features \cite{r6, r24, r25, r26, r34}. Similarly, zigzag persistence provides a flexible framework to track dynamic topological features across time but cannot be directly applied to sequences of grayscale images without preprocessing steps such as binarization or conversion to point clouds to extract meaningful topological structure \cite{r8, r2}. Crocker plots, while capable of tracking the number of topological features over time, lack the ability to distinguish individual feature trajectories \cite{r11}. Another possible option is multiparameter persistence which allows for tracking changes in topological features across multiple scales and dimensions, such as spatial and temporal scales \cite{r23}. An extension of multiparameter persistence to a time series of images could potentially identify changes to a hole structure over time; however, this approach remains under development and poses significant computational and theoretical challenges. The field is still working toward computationally efficient, robust, and widely applicable techniques that can handle complex data, such as time series of images. 

The challenges described above in analyzing a general time series of images using TDA can be seen in our data application: tracking structural changes to a cell's wound over time. Wound healing is a dynamic process by which a protein ring forms around the wound site and gradually shrinks as the cell heals. Several topological techniques exist for tracking the feature that represents the wound over time, and the feature's {\em persistence} can be used to quantify the `size' or `prominence' of that feature. In \cite{r1}, a method was developed to estimate the persistence and uncertainty of the cell wound in an image at a single time point, and by using domain knowledge, these topological features (loops) could be connected across time, assuming the continuity of the wound. In contrast, \cite{r2} and \cite{r27} explicitly incorporate temporal dynamics by tracking specific topological features over time through their persistence. Each of these techniques provide information limited to changes in the persistence of individual loops over time, under the assumption that loops remain non-overlapping in their persistence values. However, these methods do not offer information about how loops interact over time, such as splitting, merging, or intersecting. Building on existing methodology like zigzag persistence, we propose a new approach that captures changes in the wound topology as it evolves. 

To address this goal of tracking the temporal evolution of a topological feature such as a cell wound, we propose a new approach for analyzing a time series of grayscale images as a three-dimensional dataset (i.e., the two-dimensional image, time, and the pixel intensity value in the spatiotemporal grid). We treat time as an intrinsic dimension of the data (rather than a dimension of the topological summaries of the data), allowing us to identify high-dimensional topological features (e.g., loops connected through time) directly from the image stack. We introduce a novel hypothesis testing framework to determine which of these features are statistically significant and to determine a threshold for the data (rather than using a predetermined threshold). This approach produces a binary image stack that preserves meaningful spatial and temporal structure and is grounded in statistical theory. Tracking multiple topological features across time is important in the context of TDA because it addresses a key limitation in many existing approaches: loss of a unique label or index of a feature over time. For instance, when moving from the data space to the topological summary statistic, the spatial or temporal correspondence between features is lost. As a result, while many TDA methods can detect when loops appear or disappear, they often cannot determine whether a particular loop continues, moves, splits, or merges across time points. Similarly, tracking multiple topological features across time is important in the context of the cell wound images, since this will give more insight into the pattern of the wound outside of persistence.

The remainder of the paper is structured as follows: Section~\ref{sec:introTimeSeries} includes a mathematical description of the problem, an introduction to TDA, a description of current methods in TDA applied to time series data including zigzag persistence, and a brief description of the data application. In Section~\ref{sec:Method}, the method is presented in four steps, where each subsection describes a step and includes a brief example. Section~\ref{sec:simulations} evaluates the performance of the new method compared to the performance of the method from \cite{r2} with several numerical examples. And finally, Section~\ref{sec:Data} illustrates the effectiveness of the method on cell biology images.

\section{Background}\label{sec:introTimeSeries}

\subsection{Setup}

A single image can be defined as a noisy function discretized onto a 2D grid $\mathcal{G} \subset \mathbb{Z}^2$ of size $d_1 \times d_2$, where each $(x,y)$ coordinate represents the center of the grid cells with rows and columns denoted by $x=\{1, 2, \ldots, d_1\}$ and $y=\{1, 2, \ldots, d_2\}$. Generalizing this notation to higher dimensions, let $\mathcal{A}^{\sigma}$ be an array of dimension $d_1 \times \ldots \times d_{M}$, defined as
\begin{align}
 \mathcal{A}^{\sigma} = \{f(x_1,x_2, \ldots, x_M)+\varepsilon(x_1,x_2, \ldots, x_M) : (x_1,x_2, \ldots, x_M) \in \mathcal{G}\},
\end{align}
where $\mathcal{G} \subset \mathbb{Z}^M $ is an $M$-dimensional grid of size $d_1 \times \ldots \times d_M$, with each tuple $(x_1,x_2, \ldots, x_M)$ representing a coordinate on that grid (i.e., the center of a grid cell). The function $f(x_1,x_2, \ldots, x_M)$ and the noise $\varepsilon(x_1,x_2, \ldots, x_M)$ are also defined in $M$ dimensions, and the noise is assumed to follow some symmetric distribution $\mathbf{F}(0, \sigma^2(x_1,x_2, \ldots, x_M))$ with mean $0$ and variance $\sigma^2(x_1,x_2, \ldots, x_M)$. The noise-free array, from which we aim to identify the topological features, is defined as:
\begin{align}
 \mathcal{A}^0 = \{f(x_1,x_2, \ldots, x_M): (x_1,x_2, \ldots, x_M) \in \mathcal{G} \}.
\end{align}

\begin{figure}
\centering
 \centering
 \includegraphics[width=0.8\linewidth]{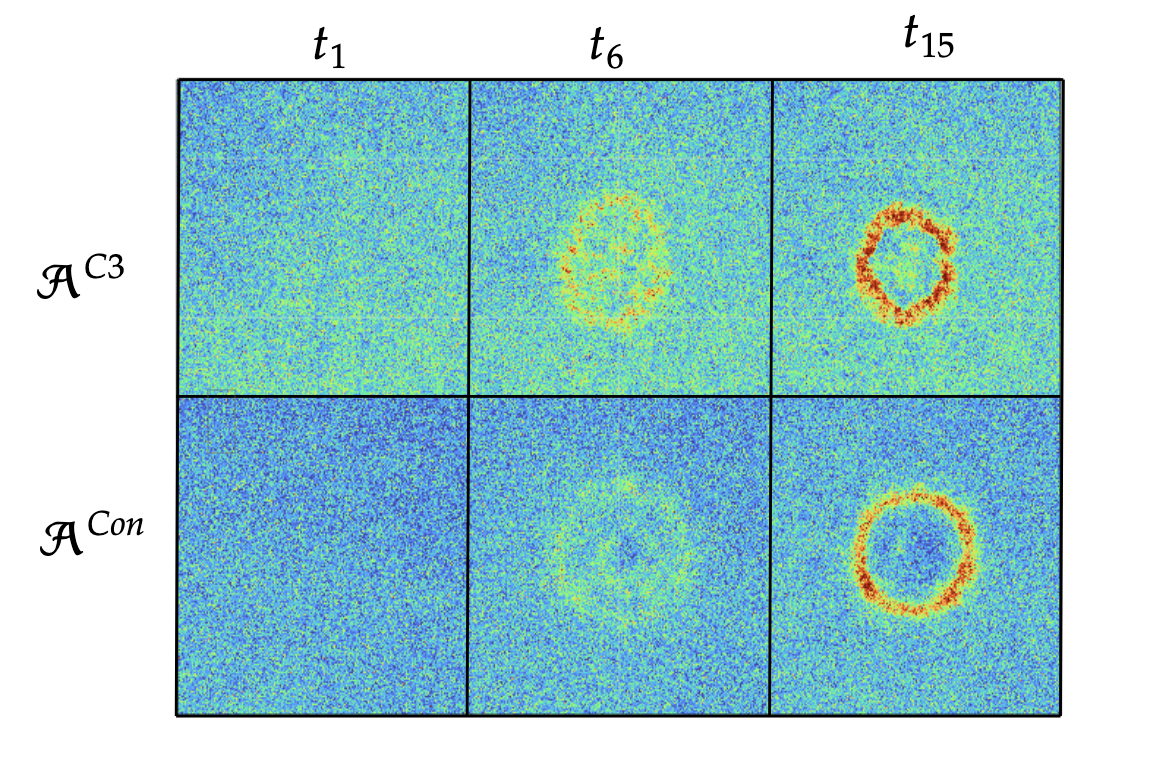}
\caption{Examples of cell wounds at different points of time in the healing process. Each row is a different cell ($\mathcal{A}^{\text{C3}}$ for the C3 cell and $\mathcal{A}^{\text{Con}}$ for the Control cell) and each column is a different time in the sequence $(t_1, t_6, t_{15})$.} \label{fig:ExampleData}
\end{figure}

The method developed in this paper is applied to a time series of images of a cell healing from the initial wound. In our data, we consider two wounded cells: the C3 cell is wounded and injected with a toxin whereas the Control cell is simply wounded. A sequence of images capture the evolution in these wounds every eight seconds. The start time for the wounding process is $t_1=0$ seconds and the last image is at $t_{30}=240$ seconds. Example images of the cell wound for both cells (C3 and Control) at distinct points in time $\{t_1, t_{6}, t_{15}\}$ in the time series of images are shown in Figure~\ref{fig:ExampleData}. 

\subsection{TDA Background}
Persistent homology (PH) is a popular tool from TDA which can be use to characterize static, and by extension time-varying, topological features in data. Each $m$-dimensional hole in an image $\mathcal{A}^{\sigma}$ is included in a homology group ($H_m(\mathcal{A}^{\sigma})$). The rank of this group is denoted by its Betti number, $\beta_m$, which can be thought of as the number of $m$-dimensional holes in the group. Examples of different dimensional topological features are zero-dimensional homology group generators ($H_0(\mathcal{A}^{\sigma})$) which represent the connected components in $\mathcal{A}^{\sigma}$, one-dimensional homology group generators ($H_1(\mathcal{A}^{\sigma})$) which represent the loops, two-dimensional homology group generators ($H_2(\mathcal{A}^{\sigma})$) which represent voids, and higher-dimensional homology group generators ($H_m(\mathcal{A}^{\sigma})$ for $ m > 2$). The term ``topological feature'' or “feature” refers to a homology group generator. When considering the homology of an image as a whole, all pixels are neighboring each other and represent a fully connected space, making the homology uninformative. In order to identify meaningful topological features in an image, which is a noisy function, we need a more sophisticated approach to estimating the homology.

PH captures changes in the hole structure of a space at multiple scales through a {\em filtration} defined using a threshold value $\delta$; the homology is tracked as $\delta$ changes. Standard choices of filtrations for images are based on upper-level or lower-level sets; we focus on upper-level sets in this paper defined as follows:
\begin{equation}
 \mathcal{A}^{[\delta, \infty)} = \{Z(x_1,x_2,\ldots,x_M) \geq \delta: (x_1,x_2,\ldots,x_M) \in \mathcal{G} \} \text{ (upper-level), }
\end{equation} 
where $Z(x_1,x_2,\ldots,x_M)=f(x_1,x_2,\ldots,x_M)+\varepsilon(x_1,x_2,\ldots,x_M)$ represents the intensity value of a pixel located at the $(x_1,x_2,\ldots,x_M)$ coordinate in the $M$-dimensional grid. The $\delta$ value in which a homological feature $j$ first appears is its `birth' time (denoted $b_j$) and the $\delta$ value in which the feature disappears or merges with other features is its `death' time (denoted $d_j$). The `lifetime' of that feature is its persistence ($p_j = b_j-d_j$); one interpretation is that higher persistence indicates a feature is topological signal and lower persistence indicates a feature is topological noise \cite{r18}. A `persistence diagram' is a plot of the birth versus death times of the features across the filtration for a given dataset. 

\begin{figure}
\centering
 \begin{subfigure}{.25\linewidth}
 \centering
 \includegraphics[width=\linewidth]{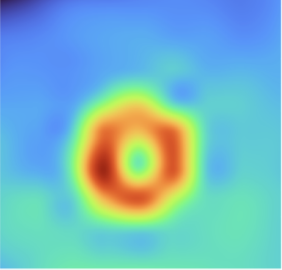}
 \caption{$\mathcal{\tilde A}^{\text{C3}}_{t_{15}}$}
 \label{subfig:ExampleSmooth}
 \end{subfigure}
 \begin{subfigure}{.25\linewidth}
 \centering
 \includegraphics[width=\linewidth]{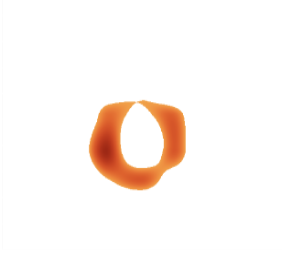}
 \caption{$b_j = 2007$}
 \label{subfig:ExampleBirth}
 \end{subfigure}
 \begin{subfigure}{.25\linewidth}
 \centering
 \includegraphics[width=\linewidth]{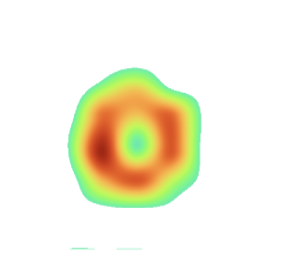}
 \caption{$d_j = 1404$}
 \label{subfig:ExampleDeath}
 \end{subfigure}
 \begin{subfigure}{.05\linewidth}
 \centering
 \includegraphics[width=\linewidth]{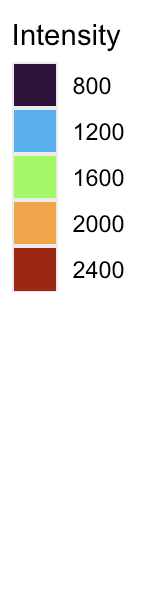}
 \label{subfig:legend}
 \end{subfigure}
 \begin{subfigure}{.5\linewidth}
 \centering
 \includegraphics[width=\linewidth]{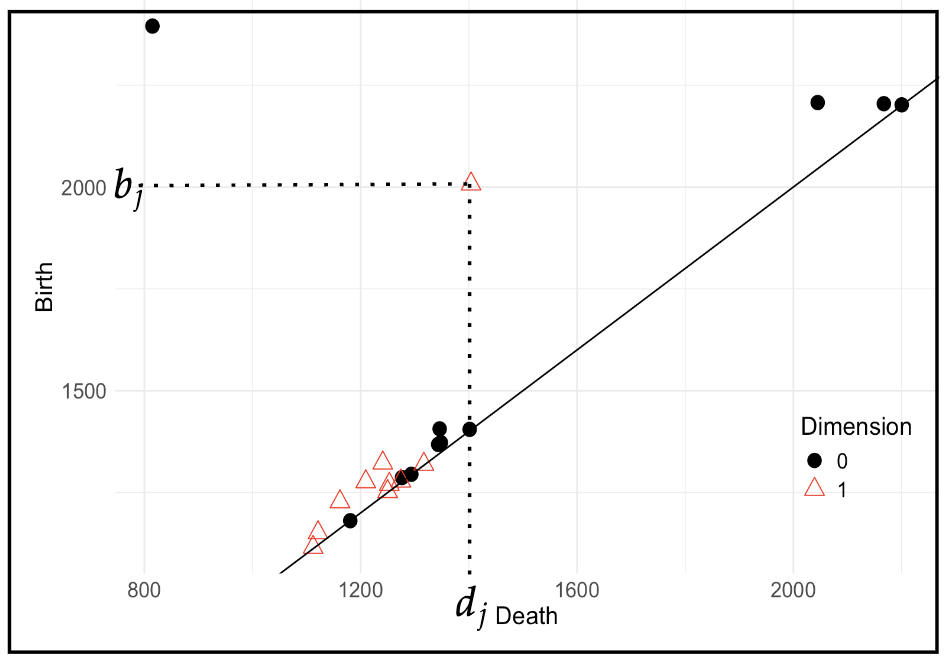}
 \caption{$\mathcal{P}(\mathcal{\tilde A}^{\text{C3}}_{t_{15}})$}
 \label{subfig:ExamplePD}
 \end{subfigure}
\caption{(a) Smoothed C3 cell at time $t_{15}$, denoted as $\mathcal{\tilde A}^{\text{C3}}_{t_{15}}$. (b) The upper-level set $(\mathcal{\tilde A}^{\text{C3}}_{t_{15}})^{[b_j, \infty)}$ at the birth of the most persistent loop with the birth time $b_j=2007$. (c) The upper-level set at the death of the most persistent loop $(\mathcal{\tilde A}^{\text{C3}}_{t_{15}})^{[d_j, \infty)}$ with the death time $d_j=1404$. (d) The persistence diagram of the upper-level set filtration of $\mathcal{\tilde A}^{\text{C3}}_{t_{15}}$ where the black dots represent connected components and the red triangles represent loops. The birth time is on the $y-$axis and the death time is one the $x-$axis. The most persistent loop is shown at coordinate ($d_j, b_j$).} \label{fig:ExampleHomology}
\end{figure}

An example of an upper-level set filtration of the smoothed image of the cell wound for the C3 cell at time $t_{15}$ can be seen in Figure~\ref{fig:ExampleHomology}: the top row shows the full image, the upper-level set at the birth time of the most persistent loop, and the upper-level set of the death time of the most persistent loop. We used a smoothed image since the cell images in this data application due to the noise in the images. For reference, the number of loops detected using an upper-level set filtration on the unsmoothed C3 cell image at time $t_{15}$ is $\beta_1=39,895$ ($\mathcal{A}^{\text{C3}}_{t_{15}}$ is in the first row and third column of Figure~\ref{fig:ExampleData}), while the true number of loops is one (due to the wounded cell). Since we are only interested in the wound, the image can be smoothed using local polynomial regression with degree 2 and span 0.1 as seen in Figure~\ref{subfig:ExampleSmooth}. The span defines the number of data points used to fit the model for each pixel (e.g., a span of 0.1 indicates that $10\%$ of the nearest pixels are used). The number of loops detected when using an upper-level set filtration on the smoothed image ($\mathcal{\tilde A}^{\text{C3}}_{t_{15}}$) is $\beta_1=10$ with the most persistent loop representing the wound. 

To facilitate persistent homology computation, we assume that the underlying space is triangulable, allowing for the approximation of the homology of the image using a simplicial complex $\mathcal{K}$. As part of our method we use a technique called zigzag persistence, introduced in Section~\ref{subsec:ZZ}. Zigzag persistence was originally developed for simplicial complexes rather than cubical complexes, which are more commonly used in TDA for image data \cite{r14}. Consequently, most existing papers and software implementations of zigzag persistence are designed for simplicial complexes. For this reason, we apply simplicial homology to our image data \cite{r20, r14}. A simplicial complex is made up of a finite set of simplexes (vertices, edges, triangles, tetrahedra, etc.) which follow the conditions: \\
(i) If a simplex $\sigma_0 \in \mathcal{K}$, then all of its faces $\tau$ must also belong to $\mathcal{K}$. The face $\tau$ of a simplex $\sigma_0$ is the convex hull of any non-empty subset of the vertices of $\sigma_0$. Each face is itself a simplex. \\
(ii) For any two simplexes $\sigma_0, \sigma_1 \in \mathcal{K}$, the intersection is either $\sigma_0 \cap \sigma_1 \in \mathcal{K}$ or $\sigma_0 \cap \sigma_1 = \emptyset$ \cite{r3}. \\
The sequence of nested sub-complexes $\mathcal{K}_\delta$ is a filtration built from the upper-level sets of $\mathcal{A}^{[\delta, \infty)}$ as $\delta$ decreases from $\infty$ to $0$ such that:
\begin{align}\label{SCTraditional}
 \mathcal{K}_{\delta_1} \xhookrightarrow{} \mathcal{K}_{\delta_2} \xhookrightarrow{} \ldots \xhookrightarrow{} \mathcal{K}_{\delta_l=0},
\end{align}
where the inclusion arrow $\xhookrightarrow{}$ indicates that $\mathcal{K}_{\delta_1} \subset \mathcal{K}_{\delta_2}$ for $\delta_1 \geq \delta_2$.

The inclusion relations between simplicial complexes induce linear maps between the homology groups of those complexes where all the linear maps point in the same direction.
\begin{align}\label{HomologyTranditional}
 H_m(\mathcal{K}_{\delta_1}) \rightarrow H_m(\mathcal{K}_{\delta_2}) \rightarrow \ldots \rightarrow{} H_m(\mathcal{K}_{\delta_l=0}).
\end{align}
This representation of vector spaces $H_m(\mathcal{K})$ is called a {\em persistence module} \citep{r14}.

The mathematical relationship shown above allows for clear interpretation, calculation, and a multi-scale view of homology over a dataset. The birth and death times of the homology groups can be represented on a topological summary statistic called a {\em persistence diagram}. An example of a persistence diagram can be seen in Figure~\ref{subfig:ExamplePD} where the detected $H_1$ features in the upper-level set filtration are the red triangles and the $H_0$ features are the black dots. The birth and death times are shown on the $y-$axis and $x-$axis, respectively. The persistence of a feature can be visualized as its relative position to the diagonal line (birth=death) where far away features are more persistent than features near the line. However, this diagram summarizes the topology of static data since there is no temporal information unless the filtration is through time. Even if the filtration is through time, topological features can disappear and reappear, losing the nesting structure needed to calculate the birth and death times of a feature.

\subsection{Zigzag persistence}\label{subsec:ZZ}
One approach which describes the PH of a sequence of spaces without requiring nesting in the filtration is zigzag persistence \citep{r14, r15, r16, r17, r8}. This technique is commonly applied to time series data where the filtration is through time, not space or function values (e.g., pixel intensity). Zigzag persistence uses ideas from representation theory of quivers which are flexible directed graphs that allow for loops and multi-arrows. Representations of quivers assign a vector space to each vertex and a linear map (homomorphism) to each arrow. The direction of each inclusion map in Equation~\eqref{SCTraditional} and the linear map in Equation~\eqref{HomologyTranditional} is arbitrary as opposed to a standard filtration where the direction is the same throughout. In many cases the inclusion direction alternates earning the name zigzag \cite{r14}. An example of this structure for a filtration over time, where $t_o$ denotes time and $o \in \{1, \ldots, l \}$, is as follows:

\begin{equation}
\begin{split}
 \mathcal{K}_{t_1} &\xhookrightarrow{} \mathcal{K}_{t_2} \xhookleftarrow{} \mathcal{K}_{t_3} \xhookrightarrow{} \ldots \xhookleftarrow{} \mathcal{K}_{t_{l-1}} \xhookrightarrow{} \mathcal{K}_{t_l} \\
 \text{or} & \\
 \mathcal{K}_{t_1} &\xhookleftarrow{} \mathcal{K}_{t_2} \xhookrightarrow{} \mathcal{K}_{t_3} \xhookleftarrow{} \ldots \xhookrightarrow{} \mathcal{K}_{t_{l-1}} \xhookleftarrow{} \mathcal{K}_{t_l}.
\end{split}
\end{equation}

The inclusion map induces a linear map on the $mth$ homology group which can now change direction creating new persistence modules:
\begin{equation}
\begin{split}\label{eq:ZZpersistencemodule}
 H_m(\mathcal{K}_{t_1}) &\rightarrow H_m(\mathcal{K}_{t_2}) \leftarrow H_m(\mathcal{K}_{t_3}) \rightarrow \ldots \leftarrow H_m(\mathcal{K}_{t_{l-1}}) \rightarrow H_m(\mathcal{K}_{t_l}) \\
 \text{or} & \\
 H_m(\mathcal{K}_{t_1}) &\leftarrow H_m(\mathcal{K}_{t_2}) \rightarrow H_m(\mathcal{K}_{t_3}) \leftarrow \ldots \rightarrow H_m(\mathcal{K}_{t_{l-1}}) \leftarrow H_m(\mathcal{K}_{t_l}).
\end{split}
\end{equation}

Note that zigzag persistence is defined for a temporal set of simplicial complexes rather than images. To apply it to our time series data, we must first convert the images into simplicial complexes that represent the relevant topological features. A detailed explanation of this transformation is provided in Section~\ref{subsec:step4}. In the next section, we propose a method to estimate topological features in $\mathcal{A}^0$ as they change in time using zigzag persistence. 

\section{Method}\label{sec:Method}
The proposed method for detecting loops in a time series of images is called the {\em Maximum Void} ({\bf MV}) method, and has four main steps. Each step is summarized below with its connection to the cell wound data highlighted. More detailed explanations are provided in the subsections that follow.\\
\textbf{Step 1:} Add a temporal dimension to the $(M-1)-$dimensional arrays in order to identify higher dimensional homology groups: $H_{M-1}$ features. For the cell wound application add a temporal dimension to the two-dimensional images in order to identify $H_2$ features.\\
\textbf{Step 2:} Identify a statistically significant $H_{M-1}$ feature which represents changes to lower-dimensional topological features throughout time. For the cell wound application find a statistically significant $H_2$ feature which represents changes to the wound ($H_1$ feature) throughout time.\\
\textbf{Step 3:} Use the birth time of the statistically significant $H_{M-1}$ feature to identify lower dimensional features $(H_0, H_1, \ldots, H_{M-2})$ which are a part of the time series. For the cell wound application use the birth time of the $H_2$ feature which represents the wound to threshold the images creating a upper-level sets that form a partition of the wound at each point in time and result in $H_0$ and $H_1$ features.\\
\textbf{Step 4:} Connect the lower dimensional features $(H_0, H_1, \ldots, H_{M-2})$ at consecutive time points and summarize the changes in homology across time using zigzag persistence. For the cell wound application connect the $H_0$ and $H_1$ features which make up wound at each point in time with zigzag persistence.

\subsection{Step 1: Add Time Dimension}\label{subsec:step1}
Focusing on the set of images from time $t_1$ to time $t_l$ ($\{ \mathcal{A}^0_{t_1}, \ldots, \mathcal{A}^{0}_{t_l} \}$), the noise free array $\mathcal{A}^0=\{f(x_1,\ldots,x_{M-1},t_o): (x_1,\ldots,x_{M-1},t_o) \in \mathcal{G} \}$ is defined by a function $f(x_1,\ldots,x_{M-1},t_o)$ discretized onto a $M-$dimensional grid $\mathcal{G}$ of size $d_1 \times \ldots \times d_M$. For example, 2D images are defined as $\mathcal{A}^0=\{f(x,y,t_o): (x,y,t_o) \in \mathcal{G} \}$ with $(x,y,t_o)$ coordinates for the rows $x$, columns $y$, and times $t_o$ in a 3D grid $\mathcal{G}$. The data, $\mathcal{A}^{\sigma}$, is defined as follows:
\begin{align}
 \mathcal{A}^{\sigma}=\{f(x_1,\ldots,x_{M-1},t_o) + \varepsilon(x_1,\ldots,x_{M-1},t_o) :(x_1,\ldots,x_{M-1},t_o) \in \mathcal{G}\} \},
\end{align}
where $f(x_1,\ldots,x_{M-1},t_o)$ is the mean for the pixel intensity in the $M-$dimensional space location $(x_1,\ldots,x_{M-1},t_o)$. Each pixel intensity $Z(x_1,\ldots,x_{M-1},t_o)$ in the array is drawn from the following distribution:
\begin{align}
 Z(x_1,\ldots,x_{M-1},t_o) \sim \mathbf{F}(f(x_1,\ldots,x_{M-1},t_o), \sigma^2(x_1,\ldots,x_{M-1},t_o)).
\end{align}

During an upper-level set filtration on $\mathcal{A}^{\sigma}$, $H_0$ through $H_{M-2}$ features may be detected. These features may arise from the true signal $f(x_1,\ldots,x_{M-1},t_o)$ or from noise $\varepsilon(x_1,\ldots,x_{M-1},t_o)$. In this setup, the $H_{M-1}$ features represent the changes through time. We assume that the function $f(x_1,\ldots,x_{M-1},t_o)$, which forms the mean of the pixel intensity sampling distribution and represents the true pattern of interest, can be partitioned as described in the following assumption.

\begin{assumption}\label{as:FunctionTime}
For each point in time $t_o \in \{t_1, \ldots, t_l\}$ an image $\mathcal{A}^{0}_{t_o}$ in the array $\mathcal{A}^{0}$ can be partitioned into $k$ contiguous regions, where $k$ could depend on time (i.e., $k_{t_o}$), and each of the $k$ partitions has a constant function value defined as:
\begin{align}
\begin{split}
 \theta_k(t_o)=\{f(x_1,\ldots,x_{M-1},t_o):(x_1,\ldots,x_{M-1}) \in \mathcal{G}_k(t_o) \\ \text{ and } f(x_1,\ldots,x_{M-1}) = \mu_k \},
\end{split}
\end{align}
where $\mathcal{G}_k(t_o)$ is the $(x_1,\ldots,x_{M-1})$ coordinates which make up partition $k$ at time $t_o$ and $\mu_k$ is the value of the function for partition $k$. Each partition represents a topological feature $\gamma_k \in H_m(G_k(t_o))$ for $m = 0,1, \ldots, M-2$.
\end{assumption}

For this paper, we assume that some partition of $\mathcal{G}$, denoted as $\mathcal{G}_1(\boldsymbol{t^*})= \{\mathcal{G}_1(t_o) : \forall t_o \in \boldsymbol{t^*} \} \subset \mathcal{G}_{d_1 \times \ldots \times d_M}$, contains a feature of interest $\gamma_1 \in H_m(\mathcal{G}_1(\boldsymbol{t^*}))$ where $\boldsymbol{t^*}$ is an ordered contiguous subset of $\{t_1, \ldots, t_l\}$; for the cell wound application $\gamma_1$ is the wound in time such that $\gamma_1 \in H_2(\mathcal{G}_1(\boldsymbol{t^*}))$. Then $\forall t_o \in \boldsymbol{t^*}$ lower-dimensional features are formed in the partitions $\mathcal{G}_1(t_o)$ at each time point where $\gamma_1$ exists and correspond to functional values $\theta_1(t_o)$. For a time series of 2D images $\gamma_1$ is a cavity such that each time slice of the partition defined by $\gamma_1$ contains some combination of loops and connected components. While the functional values remain constant within each time slice, the overall set of values that define $\gamma_1$ may change over time.

\begin{figure}
\centering
 \includegraphics[width=1\linewidth]{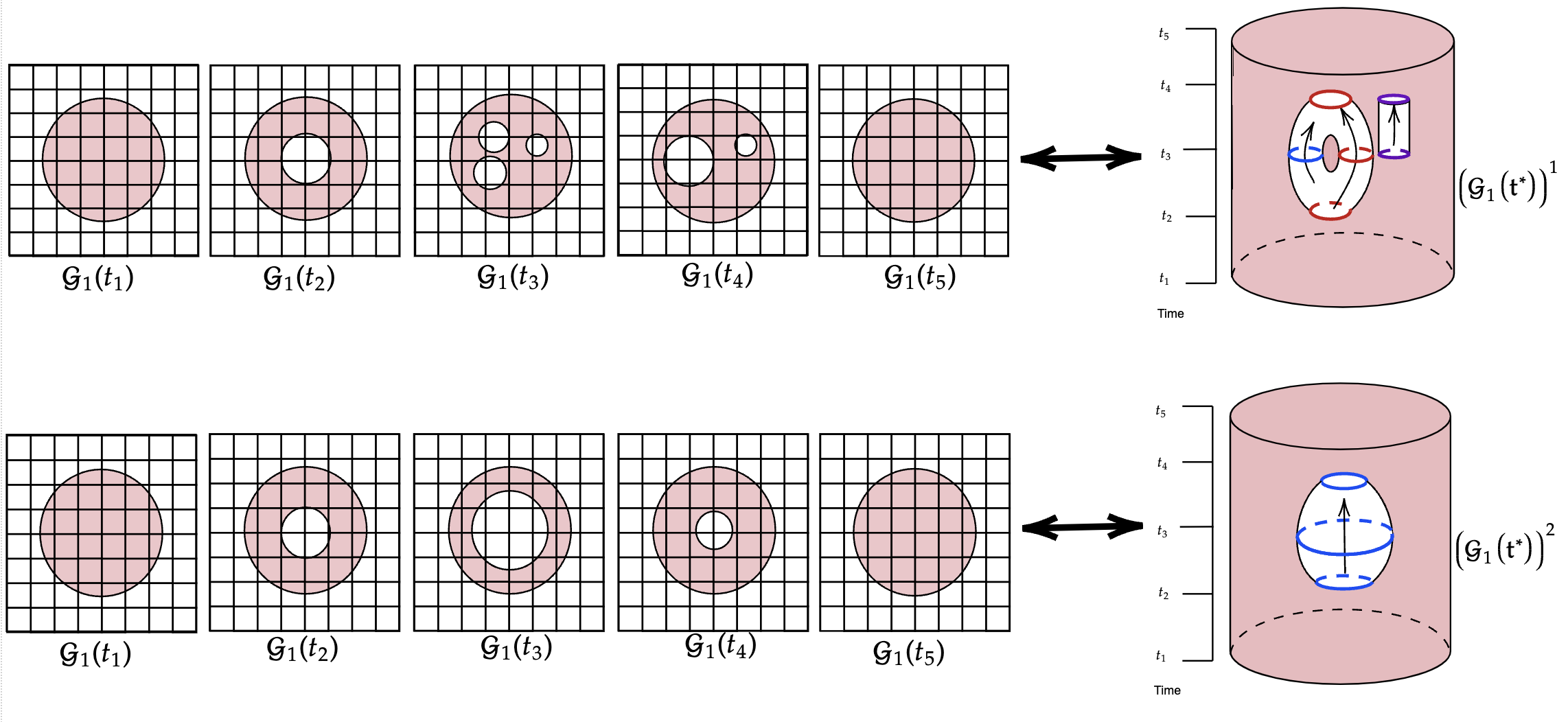}
 \caption{Each row is a partition of a topological feature in an image set changing across time. The left-side graphics are the partitions of loops and connected components at each point in time $t_o \in \{ t_1, \ldots, t_5 \}$ for two different time series of images: $(\mathcal{A}^{\sigma}_{t_o})^1$ (row 1) and $(\mathcal{A}^{\sigma}_{t_o})^2$ (row 2). The right-side graphics are the corresponding higher-order partition of the topological feature across time $(\mathcal{G}_1(\boldsymbol{t^*}))^1$ (row 1) and $(\mathcal{G}_1(\boldsymbol{t^*}))^2$ (row 2) where the arrows and the colors symbolize changes in a specific loop across time. }\label{fig:ExampleTimeSeries}
\end{figure}

Figure~\ref{fig:ExampleTimeSeries} illustrates examples of how lower-dimensional features can evolve in a time series of 2D images and how these changes can be characterized by higher-dimensional features. Two time series of images are presented in which the topological feature at each time is the white interior of the pink partition, denoted $(\mathcal{G}_1(t_o))^1$ and $(\mathcal{G}_1(t_o))^2$. The evolution of the holes across consecutive time-point may be difficult to understand when viewing the partitions side by side. The cavities which are also the white interiors of the pink partitions, denoted $(\mathcal{G}_1(\boldsymbol{t^*}))^1$ and $(\mathcal{G}_1(\boldsymbol{t^*}))^2$ where $\boldsymbol{t^*} = \{t_1, \ldots, t_5\}$, display the interpretation of the loops' evolutions with colors and arrows. In particular, $(\mathcal{G}_1(\boldsymbol{t^*}))^1$ which has a similar structure to the C3 cell is tracking a more complex temporal pattern development due to multiple holes, compared to $(\mathcal{G}_1(\boldsymbol{t^*}))^2$ which has a similar structure to the Control cell with only one hole. For $(\mathcal{G}_1(\boldsymbol{t^*}))^1$, $\gamma_1$ can be the red cavity or the purple cavity, while for $(\mathcal{G}_1(\boldsymbol{t^*}))^2$, $\gamma_1$ is the blue cavity. In the next subsection, we outline how to identify the partition $\mathcal{G}_1(\boldsymbol{t^*})$ in $\mathcal{A}^{\sigma}$ (for the cell biology application, this is the partition corresponding to the cell wound) with a new hypothesis testing framework to identify $\gamma_1$.

\subsection{Step 2: Find Statistically Significant $H_{M-1}$ Feature}\label{subsec:step2}
Any $H_{M-1}$ features in $\mathcal{A}^{\sigma}$ may be signal or noise, so we propose a way to detect a true $H_{M-1}$ feature in order to identify the partition $\mathcal{G}_1(\boldsymbol{t^*})$ which connects partitions $\mathcal{G}_1(t_o)$ in the image set across time. The targeted $\gamma_1$ is the most persistent $H_{M-1}$ feature with birth time $\theta^*_1=\min\{f(x_1,\ldots,x_{M-1},t_o): (x_1,\ldots,x_{M-1},t_o) \in \mathcal{G}_1(\boldsymbol{t^*})\}$ which is a good threshold to identify $\mathcal{G}_1(\boldsymbol{t^*})$ from $(\mathcal{A}^0)^{[\theta_1^*, \infty)}$. This assumption is reasonable for the cell biology application though generalizations to any images with ring structures are possible. More formally, we assume the following:

\begin{assumption}\label{as:PersistenceWoundTime}
 For every time $t_o \in \boldsymbol{t^*}$ where $\gamma_1$ exists, the persistence of the corresponding $H_{M-1}$ feature is higher than the persistence of the background ($k \neq 1$) such that: $\theta_1(t_o) > \theta_k(t_o)$ if $\mathcal{G}_1(t_o)$ is an $H_0$ feature and $\theta_{1}(t_o)-\theta_{1*}(t_o) > \theta_{k}(t_o)-\theta_{k*}(t_o)$ if $\mathcal{G}_1(t_o)$ is an $H_m$ feature where $m \neq 0$; $\theta_{1*}$ and $\theta_{k*}$ denote the partitions in the interior of loop $1$ and loop $k$, respectively.
\end{assumption}

\begin{assumption}\label{as:FunctionAcrossTime}
 For every pair of times $t_i,t_j \in \boldsymbol{t^*}$, the functional value of the partition $\mathcal{G}_1(t_i)$ is closer to the functional value of the partition at another point in time $\mathcal{G}_1(t_j))$ than it is to any other partition so that: $\theta_1(t_i) - \theta_1(t_j) < \theta_1(t_i) - \theta_k(t_i)$ where $k \neq 1$ and $i \neq j$.
\end{assumption}

Under these assumptions, the proposed hypothesis test, called {\em Maximum Persistence Test}, can be used to identify $\gamma_1$.

\subsubsection{Maximum Persistence Test}\label{subsubsec:max_test}

To test if any of the features detected in the array $\mathcal{A}^{\sigma}$ are statistically significant, we apply a hypothesis test called the {\em Maximum Persistence Test} where the null hypothesis assumes there are no true features in the image. When outlining the test we use the $M$-dimensional array notation to keep the test general as it can be applied to any array not just a time series of 2D images. Specifically, assume that $f(x_1,x_2, \ldots, x_M)$ has a real feature $(H_m)_j$ of a particular homology dimension $m$ where $m \in \{0, \ldots, M-1\}$, $j = \{1, \ldots, \beta_m \}$, and $\beta_m$ is the total number of true $H_m$ features. Then the persistence of $(H_m)_j$ should be higher than most of the persistence of the $H_m$ features under the null hypothesis, $(H_m)_{\text{null}}$, where there are no true features present in the image. We focus on the maximum persistent feature (e.g., $\gamma_1$), since a common interpretation is that higher persistence is associated with a real topological signal. We assume the following two conditions for the Maximum Persistence Test under the null hypothesis.

\begin{assumption}\label{as:NullMT}
For the null hypothesis of the Maximum Persistence Test, assume that the array $\mathcal{A}^0$ has no true features (other than an $H_0$ feature) such that $\mathcal{A}_{\text{null}}^0=\{\mu_0 : \forall (x_1,x_2, \ldots, x_M) \in \mathcal{G} \}$, for a fixed $\mu_0$. This implies that there are no true $H_m$ features, $m=1, \ldots, M-1$. 
\end{assumption}
\noindent In this case, $\mu_0$ denotes the mean of the background partition where there are no features.

\begin{assumption}\label{as:HomoskedasticNoise}
\sloppy For simplicity, assume that the noise is homoskedastic in $\mathbf{F}(0, \sigma^2(x_1,x_2, \ldots, x_M))$ such that $\mathcal{A}_{\text{null}}^{\sigma} = \{ \mu_0 + \sigma_0:\forall (x_1,x_2, \ldots, x_M) \in \mathcal{G} \}$. 
\end{assumption}

Given these assumptions, any features detected in $\mathcal{A}_{\text{null}}^{\sigma}$ have pixels sampled from the following distribution:
\begin{equation}\label{eq:PixelsNull}
 Z(x_1,\ldots,x_M) \sim \mathbf{F}(\mu_0, \sigma_0^2).
\end{equation}

For a given array, an upper-level set filtration on $\mathcal{A}^{\sigma}$ and its corresponding persistence diagram $\mathcal{P}(\mathcal{A}^{\sigma})$ have $\beta_m$ $H_m$ features for each dimension $m=\{0, \ldots, M-1\}$. The $(H_m)_j$ features detected in the array have birth and death times, $\{(d_1^{(m)}, b_1^{(m)}), \ldots, (d_{\beta_m}^{(m)},b_{\beta_m}^{(m)}) \}$, where $d_j^{(m)}$ represents the death time and $b_j^{(m)}$ represents the birth time for $H_m$ feature $j=1, \ldots, \beta_m$. All of these birth and death times are considered topological noise assuming $H_{\text{null}}$ is true so that they are drawn from the same distribution $\mathbf{F}(\mu_0, \sigma_0^2)$ with the constraint that the birth time is larger than the death time as stated below:
\begin{align}
 d_j^{(m)}, b_j^{(m)} \overset{iid}{\sim} \mathbf{F}(\mu_0, \sigma_0^2) \text{ with } b_j^{(m)} > d_j^{(m)}, \quad j=1, \ldots, \beta_m. 
\end{align} 

For a given feature $(H_m)_j$, the null hypothesis is that the birth and death times of $(H_m)_j$ have the same distributions, that is:
\begin{equation}\label{eq:H0}
 H_{\text{null}} : d_j^{(m)}, b_j^{(m)} \sim \mathbf{F}(\mu_0, \sigma_0^2) \text{ with } b_j^{(m)} > d_j^{(m)}.
\end{equation}

For the Maximum Persistence Test, we first focus on the maximum persistent $H_m$ feature as this feature is the most likely to be statistically significant in terms of its persistence. The maximum persistent $H_m$ is defined as follows:
\begin{align}\label{eq:maxpersistent}
 \rho_{\max}^{(m)} =\max_{(d_j^{(m)},b_j^{(m)}) \in H_m(\mathcal{A}^{\sigma})} \{b_j^{(m)}-d_j^{(m)} \vert j = 1, \ldots, \beta_m \},
\end{align}
where $b_{\max}^{(m)}$ and $d_{\max}^{(m)}$ are the birth and death times, respectively, of the most persistent $H_m$ feature.

A one-sample permutation test is used to empirically generate the null distribution from Equation~\eqref{eq:H0} for the Maximum Persistence Test; we can use a one-sample test since the null hypothesis considers all the pixels in the image to be from the same distribution in Equation~\eqref{eq:PixelsNull}. Let $B$ denote the total number of permutations such that there are $B$ permuted arrays $(\mathcal{A}^{\sigma}_1)^*, \ldots, (\mathcal{A}^{\sigma}_B)^*$ from which to generate the null distribution. Let there be $N$ total pixels that make up $\mathcal{A}^{\sigma}$ where each pixel is indexed by $n \in \{1, \ldots , N \}$ denoted by its intensity $Z_n = Z(x_1^n, \ldots, x_M^n)$ where $(x_1^n, \ldots, x_M^n)$ is the coordinate of the $n^\text{th}$ pixel on the $M-$dimensional grid. The $q^{\text{th}}$ permuted array $(\mathcal{A}^{\sigma}_q)^*$ where $q = \{ 1, \ldots, B\}$ is a random permutation $\pi_q : \{1, \ldots, N \}\overset{\sim}{\rightarrow} \{1,\ldots, N \}$ where $\overset{\sim}{\rightarrow}$ is a bijection from the set of pixel indexes onto itself. The permutation $\pi(n)$ tells you which pixel intensity is assigned to location $n$ in the permuted image. The permuted image $Z_q^*(x_1^n,x_2^n, \ldots, x_M^n )$ assigns intensity values that used to be at location $(x_1^{\pi(n)},x_2^{\pi(n)}, \ldots, x_M^{\pi(n)} )$ to location $(x_1^n,x_2^n, \ldots, x_M^n )$ defined as follows:
\begin{align}
 Z_q^*(x_1^n,x_2^n, \ldots, x_M^n ) = Z(x_{1}^{\pi(n)},x_{2}^{\pi(n)}, \ldots, x_{M}^{\pi(n)} ).
\end{align}

For each new permutation $(\mathcal{A}^{\sigma}_q)^*$, we generate a persistence diagram, $\mathcal{P}((\mathcal{A}^{\sigma}_q)^*$), and get the persistence $\rho_{q,\max}^*$ of the maximum persistent $H_m$ feature:
\begin{align}
 \rho_{q,\max}^*(m) = \max_{(d_j^{(m)},b_j^{(m)}) \in H_m((\mathcal{A}^{\sigma}_q)^*)} \{b_j^{(m)}-d_j^{(m)} \vert j= 1, \ldots, \beta_{m,q}^* \},
\end{align}
where $\beta_{m,q}^*$ is the total number of $H_m$ features detected in an upper-level set filtration for each new permutation $(\mathcal{A}^{\sigma}_q)^*$.
The set $\boldsymbol{\rho}^*_{\textbf{max}}(m)=\{ \rho_{1,\max}^*(m), \ldots, \rho_{B,\max}^*(m)\}$ is used to estimate the null distribution. The birth and death times for the maximum persistent $H_m$ feature in permuted array $q$ are denoted as $((d_{\max}^{(m)})_q^*, (b_{\max}^{(m)})_q^*)$.

The observed test statistic for the Maximum Persistence Test is the persistence of the maximum persistent $H_m$ feature described in Equation~\eqref{eq:maxpersistent} for the observed data $\mathcal{A}^{\sigma}$ defined as: 
\begin{align}\label{eq:tobs}
 \rho_{\max}^{\text{obs}}(m)= b_{\max}^{(m)}-d_{\max}^{(m)}.
\end{align}

The permutation p-value can be then be estimated as follows:
\begin{align}
 \text{p-value}_{\text{max}}=\frac{\sum_{q=1}^{B} I(\rho^*_{q,\max}(m) \geq \rho^{\text{obs}}_{\max}(m))}{B}
\end{align}
where $I$ is an indicator function and $\alpha$ is the significance level of the hypothesis test.

If the observed maximum persistence of the $H_m$ features, $\rho_{\max}^{\text{obs}}$, is in the upper $\alpha$ percentile of the null distribution, $\boldsymbol{\rho^*}_{\textbf{max}}(m)=\{ \rho_{1, \max}^*(m), \ldots, \rho_{B, \max}^*(m) \}$, then the null hypothesis is rejected. The statistically significant $H_m$ feature is considered topological signal (i.e., it is assumed to be a topological feature in $\mathcal{A}^0$).

Recall that the birth and death times of loops detected in an upper-level set filtration on an array $\mathcal{A}^{\sigma}$ are based on the intensities of the pixels, not on geometric information such as area of the feature. One way to indirectly include geometric information in this filtration process is to smooth the array $\mathcal{\tilde A}^{\sigma}$ so that geometrically smaller features may be eliminated by the smoothing.

\begin{figure}
\centering
\includegraphics[width=0.8\linewidth]{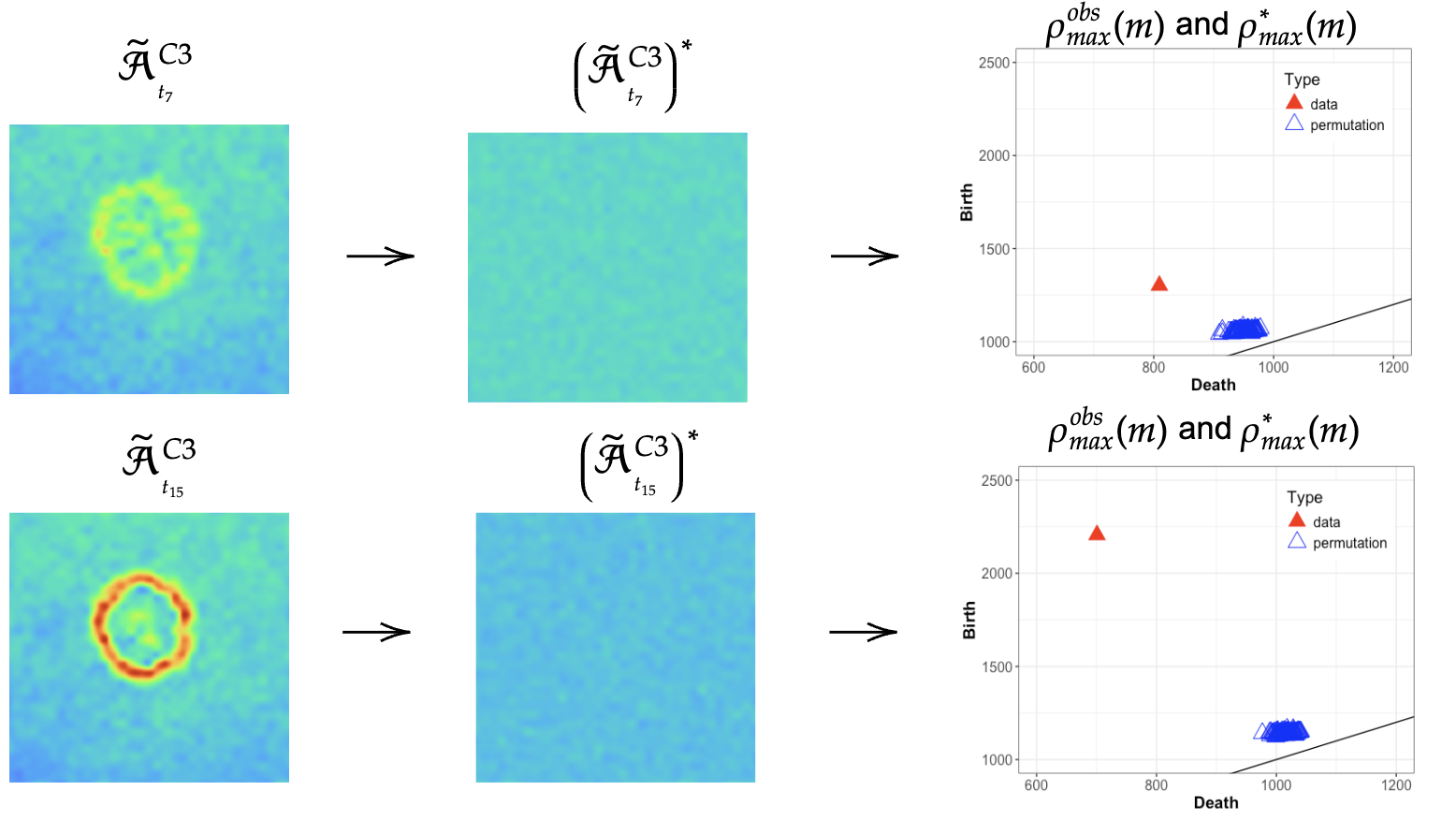}
\caption{Pipeline for the Maximum Persistence Test for $\mathcal{\tilde A}^{\text{c3}}_{t_7}$ and $\mathcal{\tilde A}^{\text{c3}}_{t_{12}}$. The first column contains the smoothed images of the wounded C3 cell at time $t_1$ (row 1) and $t_{12}$ (row 2). The second column has permuted examples of the smoothed images where the original image $\mathcal{A}^{\text{c3}}_{t}$ is permuted and then smoothed to become $(\mathcal{\tilde A}^{\text{c3}}_{t})^*$. The last column has all the maximum persistent $H_1$ features for the permutations $\boldsymbol{\rho}^*_{\textbf{max}}(m)$ (black triangles) and the maximum persistent $H_1$ feature of the data $\rho^{\text{obs}}_{\max}(m)$ (red triangle).}\label{fig:PipelineMT}
\end{figure}

An example of the pipeline for this test can be seen in Figure~\ref{fig:PipelineMT} where the first column shows the smoothed images for the C3 cell: $\mathcal{\tilde A}^{\text{C3}}_{t_7}$ (row one) and $\mathcal{\tilde A}^{\text{C3}}_{t_{12}}$ (row two). The next column shows an example of a smoothed permutation of that image, and the last column shows a persistence diagram with the maximum persistent $H_1$ feature of all the permuted images (null distribution) and the maximum persistent $H_1$ feature from the data. More details on the Maximum Persistence Test can be found in \citep{r33}. The p-values for both times are less than $0.001$ since the corresponding observed test statistics are greater than all the permuted test statistics.

\subsection{Step 3: Identify $H_0, \ldots, H_{M-2}$ Features at Each Point in Time}\label{subsec:step3}

The partition of the evolving features in $\mathcal{A}^{\sigma}$ is estimated as follows:
\begin{align}
\begin{split}
 \mathcal{\hat G}_1(\boldsymbol{t^*})= ((\mathcal{A}^{\sigma})^{[\hat \theta_1^*, \infty)}) = \{(x_1,\ldots,x_{M-1},t_o) : f^{-1}(\theta) = (x_1,\ldots,x_{M-1},t_o) \\ \text{ for } \theta \geq \hat{\theta}_1^{*} \}, 
\end{split}
\end{align}
where only $(x_1,\ldots,x_{M-1},t_o)$ coordinates from pixels above the birth time of the most persistent $H_M$ feature $\gamma_1$ are a part of the estimated partitions. This partition can be broken into slices at each point in time $t_o \in \boldsymbol{t^*}$: let the $(x_1,\ldots,x_{M-1})$ coordinates defining the topological feature at time $t_o$ be estimated as follows:
\begin{align}
\begin{split}
 \mathcal{\hat{G}}_1(t_o) = ((\mathcal{A}^{\sigma}_{t_o})^{[\hat \theta_1^*, \infty)}) = \{(x_1,\ldots,x_{M-1}) : f^{-1}(\theta) =(x_1,\ldots,x_{M-1},t_o) \\ \text{ for } \theta \geq \hat{\theta}_1^{*} \}, 
\end{split}
\end{align}

The $(x_1,\ldots,x_{M-1})$ coordinates in $\mathcal{\hat{G}}_1(t_o)$ are the zero-simplices which form the simpicial complex at each time point $\mathcal{K}_{t_o}^{\hat \theta_1^{*}}$ which can be used to calculate the homology at each point in time ($H_m(\mathcal{K}_{t_o}^{\hat \theta_1^{*}})$). The homology of the space at time $t_o$ can manifest as a lower-dimensional feature $H_0, \ldots, H_{M-1}$ feature, the empty set, or multiple features of different dimensions. 

If there are fewer changes in the topological features over time (e.g., $(\mathcal{G}_1(\boldsymbol{t^*}))^2$ in Figure~\ref{fig:ExampleTimeSeries} similar to the Control cell), then the homology at each point in time is easier to connect to the times before and after. For instance, $(\mathcal{G}_1(\boldsymbol{t^*}))^2$ would be defined by one blue loop which is a part of the wound at times $t_2, t_3,$ and $t_4$. However, if the wound is more disorganized (e.g., $(\mathcal{G}_1(\boldsymbol{t^*}))^1$ in Figure~\ref{fig:ExampleTimeSeries} similar to the C3 cell) and has multiple $H_m$ features which describe it, the homology at each point in time can be difficult to connect to the time before. For instance, $(\mathcal{G}_1(\boldsymbol{t^*}))^1$ is defined by a red loop which is a part of the wound at times $t_2,t_3,$ and $t_4$, a blue loop which is a part of the wound at time $t_3$ but then merges with the red loop at time $t_4$, and a completely separate purple loop which is a part of the wound at times $t_3$ and $t_4$. We make these connections (e.g., groupings by color) with zigzag persistence.

\subsection{Step 4: Connect the $H_1$ and $H_0$ features Across Time}\label{subsec:step4}

The final step of the {\bf MV} method involves connecting and summarizing the homology of $\mathcal{K}^{\hat \theta_1}_{t_o}$ and $\mathcal{K}^{\hat \theta_1^*}_{t_{\overline o}}$ for all the consecutive time points $t_o, t_{\overline o} \in \boldsymbol{t^*}$ through zigzag persistence. The first three steps of the method reduce the time series of images $\boldsymbol{\mathcal{\tilde A}^{\sigma}}=\{\mathcal{\tilde A}^{\sigma}_{t_{1}}, \mathcal{\tilde A}^{\sigma}_{t_{2}}, \ldots, \mathcal{\tilde A}^{\sigma}_{t_l} \}$ into a time series of simplicial complexes $\boldsymbol{\mathcal{K}^{\hat \theta_1^*}} = \{\mathcal{K}^{\hat \theta_1^*}_{t_{1}},\mathcal{K}^{\hat \theta_1^*}_{t_{2}}, \ldots, \mathcal{K}^{\hat \theta_1^*}_{t_l} \}$ (outlined in Algorithm~\ref{alg:GetH2} for 2D images). These simplicial complexes are constructed solely from pixels with intensity values above $\hat \theta_1^*$ and vary in structure over time. 

\begin{algorithm}
\begin{algorithmic}[1]
\State \textbf{Input: } $df \coloneqq(x,y,t,Z[x,y,t])$ of array $\mathcal{A}^{\sigma}$; $B=$ number of permutations; $m=$ homology group dimension; $P=$ persistence diagram of $\mathcal{A}^{\sigma}$; $r^{obs}\coloneqq \max_m(P)$ persistence of most persistent $H_m$ feature; $b=$ birth time of maximum persistent $H_m$ feature; $T$ total number of time points
\State \textbf{Output:} $S$ list of simplicial complexes for representing the wound at each point in time
\State Define: $\mathbb{Z}=\{ Z[x,y,t] \mid (x,y,t,Z[x,y,t]) \in df \}$; $p=\emptyset$; $r^*=\emptyset$; $L=$nrows$(\mathbb{Z})$

\For{q in 1:B}
 \State{Step 1:} Define $A^* \in \mathbb{R}^{3}$
 \For{l in 1:L} \Comment{permute $Z[x,y,t]$} 
 \State $A^*$(x,y,t)$ \leftarrow$ sample($\mathbb{Z}$, without replacement)
 \EndFor{}
 \State{Step 2:} $P^* \leftarrow$ pers($A^*$) 
 \Comment{calculate Persistence Diagram}
 \State{Step 3:}
 \State $r^* \leftarrow$ max$_m(P^*)$
 \Comment{Get max persistence}
 \EndFor{}
\State{Step 4:} $p \leftarrow \frac{\sum{q=1}^{B} \mathbb{I}(r^*_q > r^{obs})}{B}$
\Comment{Get p-value for Maximum Persistence Test}
\State{Step 5:}
\If{$p > \alpha$} {$S \leftarrow \emptyset$ }
\Comment{If fail to reject null there is no feature}
\Else{\State Define $S=\emptyset$; $G=\emptyset$
\Comment{If reject null find $(x,y,t)$ coordinates of feature}
\For{o in 1:T} 
\State $G_t=\leftarrow \{ (x,y,Z[x,y,t]) \mid t=o \}$
\Comment{Subset dataframe by each time}
\State $S_t \leftarrow \{ (x,y) \mid \left( Z[x,y,t] \geq b \right)\cap \left( (x,y,Z[x,y,t]) \in G_t \right) \}$
\Comment{Get $(x,y)$ for birth of feature}
\EndFor{}}
\EndIf{}
\State \Return{S}
\end{algorithmic}
\caption{\textbf{MV} method for Stacked 2D Images}\label{alg:GetH2}
\end{algorithm}

\begin{figure}
\centering
\begin{subfigure}{.45\linewidth}
 \centering
 \includegraphics[width=\linewidth]{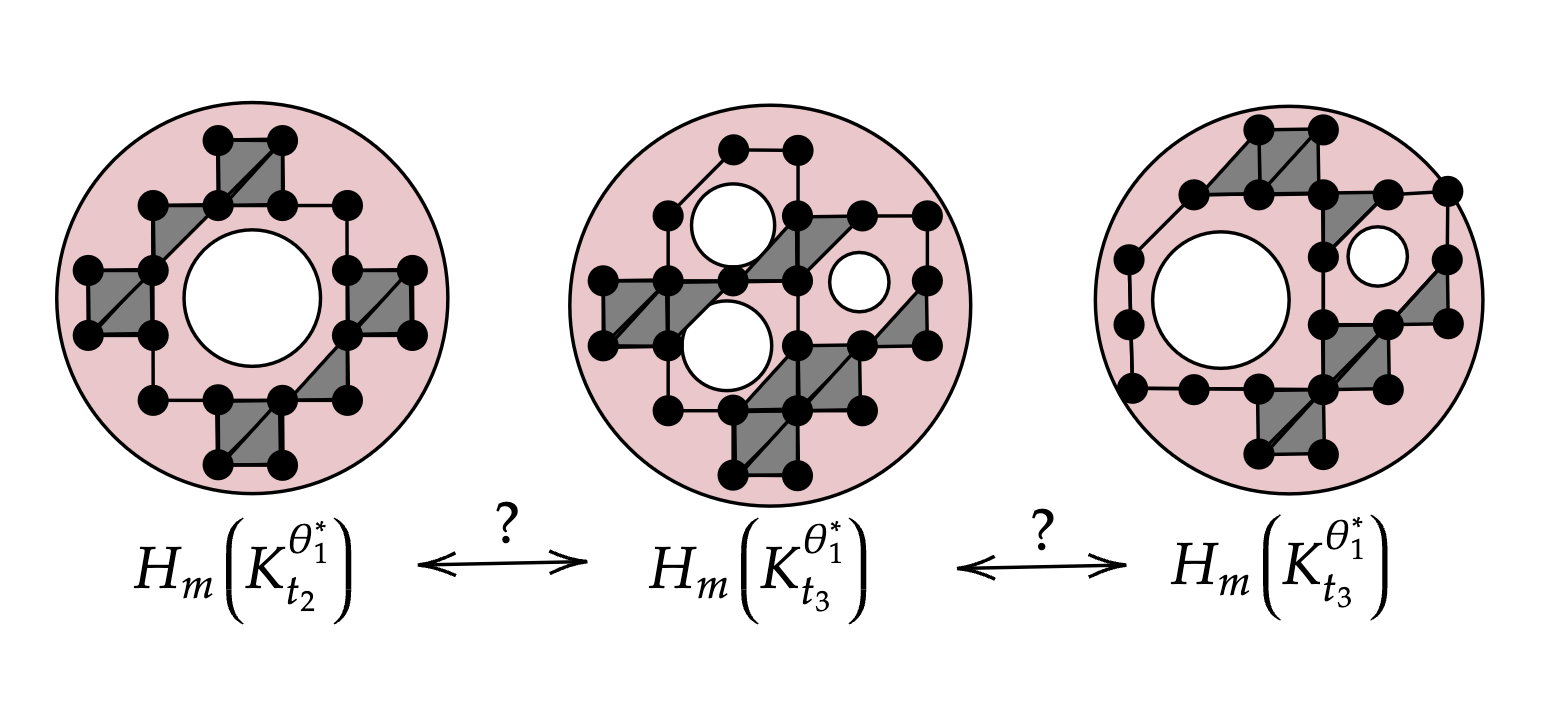}
 \caption{$\mathcal{K}_{t_o}^{\theta_1^*}$ on top of $\mathcal{G}_1(t_o)$ for $t_{2}, t_{3}, t_{4}$}\label{subfig:slices}
 \end{subfigure}
 \begin{subfigure}{.45\linewidth}
 \centering
 \includegraphics[width=\linewidth]{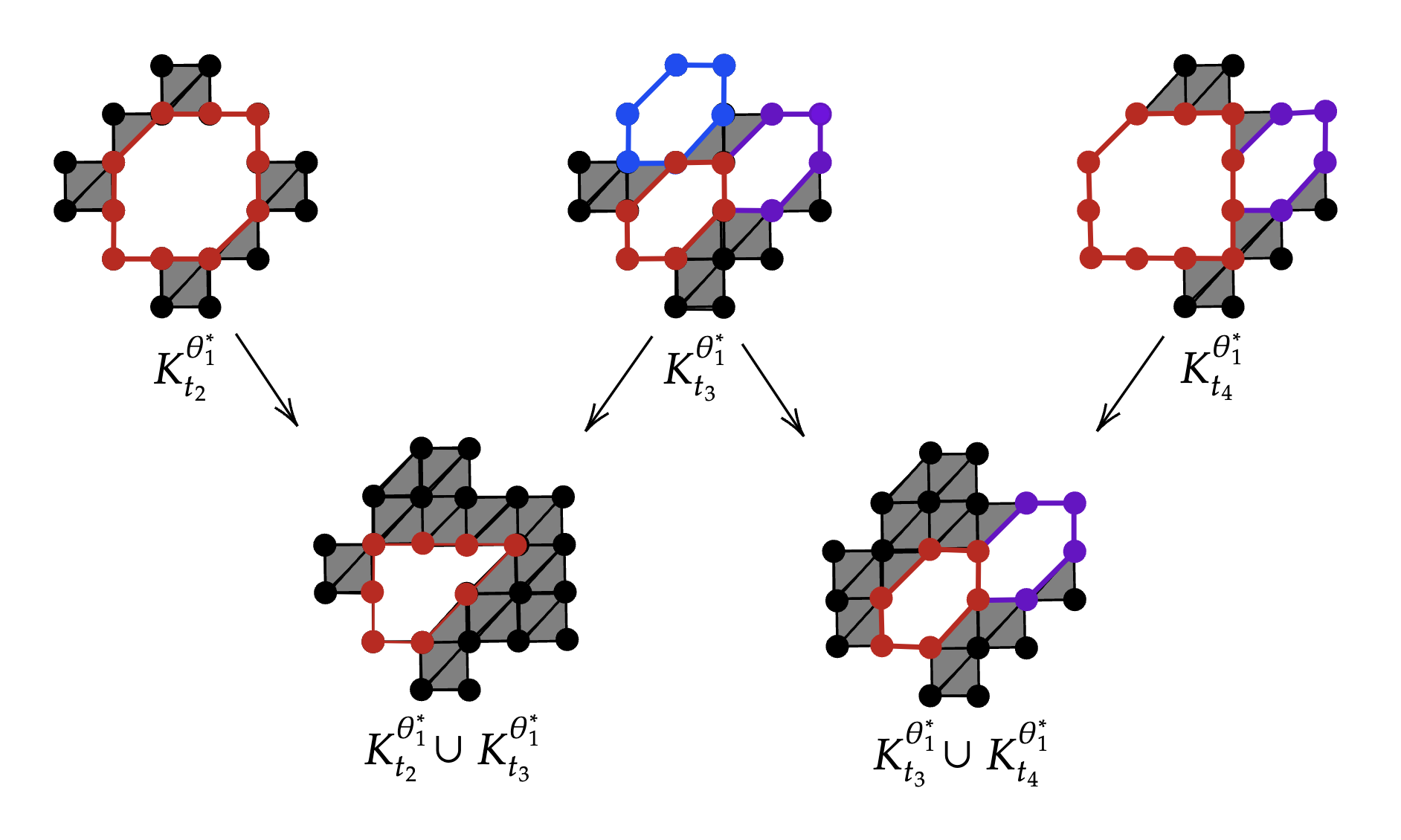}
 \caption{Zigzag of $\mathcal{K}_{t_o}^{\theta_1^*}$ for $t_{2}, t_{3}, t_{4}$}\label{subfig:sc}
 \end{subfigure}
\caption{(a) Slices of the partition $\mathcal{G}_1(t_{2})$, $\mathcal{G}_1(t_{3})$, $\mathcal{G}_1(t_{4})$. The nesting structure of the homology of the wound throughout times $t_{2}, t_{3}, t_{4}$ is unpredictable. (b) The top row of simplicial complexes are those representing the features at time points $t_{2}, t_{3}, t_{4}$ and bottom row are the unions between simplicial complexes. The loop color connects $H_1$ features through the time slices.} \label{fig:ExampleZigZag}
\end{figure}

To illustrate zigzag persistence for our cell biology application, we use the time series from the first row of Figure~\ref{fig:ExampleTimeSeries}. In Figure~\ref{fig:ExampleZigZag} we have three consecutive partitions $\{ \mathcal{G}_1(t_2), \mathcal{G}_1(t_3), \mathcal{G}_1(t_4) \}$ with their respective simplicial complexes overlaid on $\{ \mathcal{K}_{t_2}^{\hat \theta^*_1}, \mathcal{K}_{t_3}^{\hat \theta^*_1}, \mathcal{K}_{t_4}^{\hat \theta^*_1} \}$. In Figure~\ref{subfig:slices} each slice of the partition for times $\{t_2, t_3, t_4\}$ does not have the alternating nesting structure required for zigzag persistence. Each slice in time may be a loop, connected component, or empty space, allowing for linear mappings in diverse directions. Since, there is no natural zigzag order to the maps, this relationship can be induced through unions and intersections of the simplicial complexes as seen in Figure~\ref{subfig:sc}. This helps connect features throughout time which are disappearing, merging, separating, or staying the same. The resulting zigzag persistence modules for both unions and intersections with the alternating linear maps are shown below:
\begin{equation}
\begin{split}\label{eq:ZZpersistence}
 H_m(\mathcal{K}_{t_1}^{\hat \theta_1^*}) &\rightarrow H_m(\mathcal{K}_{t_1}^{\hat \theta_1^*} \cup \mathcal{K}_{t_2}^{\hat \theta_1^*}) \leftarrow H_m(\mathcal{K}_{t_2}^{\hat \theta_1^*}) \rightarrow \ldots \rightarrow H_m(\mathcal{K}_{t_l}^{\hat \theta_1^*}) \\
 \text{or} & \\
 H_m(\mathcal{K}_{t_1}^{\hat \theta_1^*}) &\leftarrow H_m(\mathcal{K}_{t_1}^{\hat \theta_1^*} \cap \mathcal{K}_{t_2}^{\hat \theta_1^*}) \rightarrow H_m(\mathcal{K}_{t_2}^{\hat \theta_1}) \leftarrow \ldots  \leftarrow H_m(\mathcal{K}_{t_l}^{\hat \theta_1^*}) 
\end{split}
\end{equation}

The choice between intersections and unions did not significantly alter the analysis; therefore, the remainder of the paper primarily focuses on unions as the set operation that connects homology over time. As seen in Figure~\ref{subfig:sc}, these set operations introduce new times points $\mathcal{K}_{t_{2}}^{\hat \theta_1^{*}} \cup \mathcal{K}_{t_{3}}^{\hat \theta_1^{*}}$ and $\mathcal{K}_{t_{3}}^{\hat \theta_1^{*}} \cup \mathcal{K}_{t_{4}}^{\hat \theta_1^{*}}$, denoted as $t_{2.5},$ and $ t_{3.5}$, which link the simplicial complexes and help interpret the evolution of the wound over the time series. The colors of the loop are defined by the unions of the simplicial complex. For instance, the red loop is the union of the loops in $\mathcal{K}_{t_{2}}^{\hat \theta_1^{*}}$ and $\mathcal{K}_{t_{3}}^{\hat \theta_1^{*}}$ and so the extra loops in $\mathcal{K}_{t_{3}}^{\hat \theta_1^{*}}$ are given a separate color.

The previous subsections present the methodological framework of the {\bf MV} method, which advances traditional TDA approaches for time series of images in two significant ways. First, it preserves the spatial and temporal continuity of topological features, and second, it incorporates statistical inference to identify and assess high-dimensional structures. We now evaluate the performance of the method using simulated data.

\section{Method Validation via Simulation}\label{sec:simulations}

\begin{figure}
\centering
\begin{subfigure}{.3\linewidth}
 \centering
 \includegraphics[width=\linewidth]{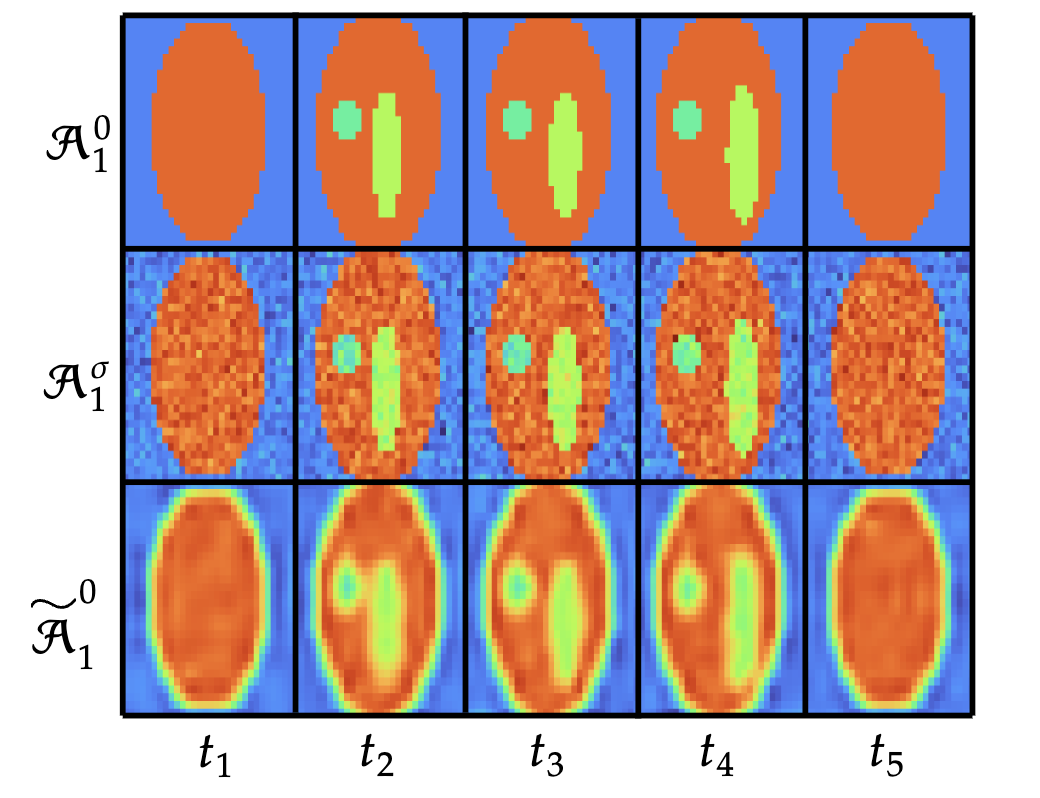}
 \caption{$\mathcal{A}_1$ example}\label{subfig:SimEx1}
 \end{subfigure}
 \begin{subfigure}{.3\linewidth}
 \centering
 \includegraphics[width=\linewidth]{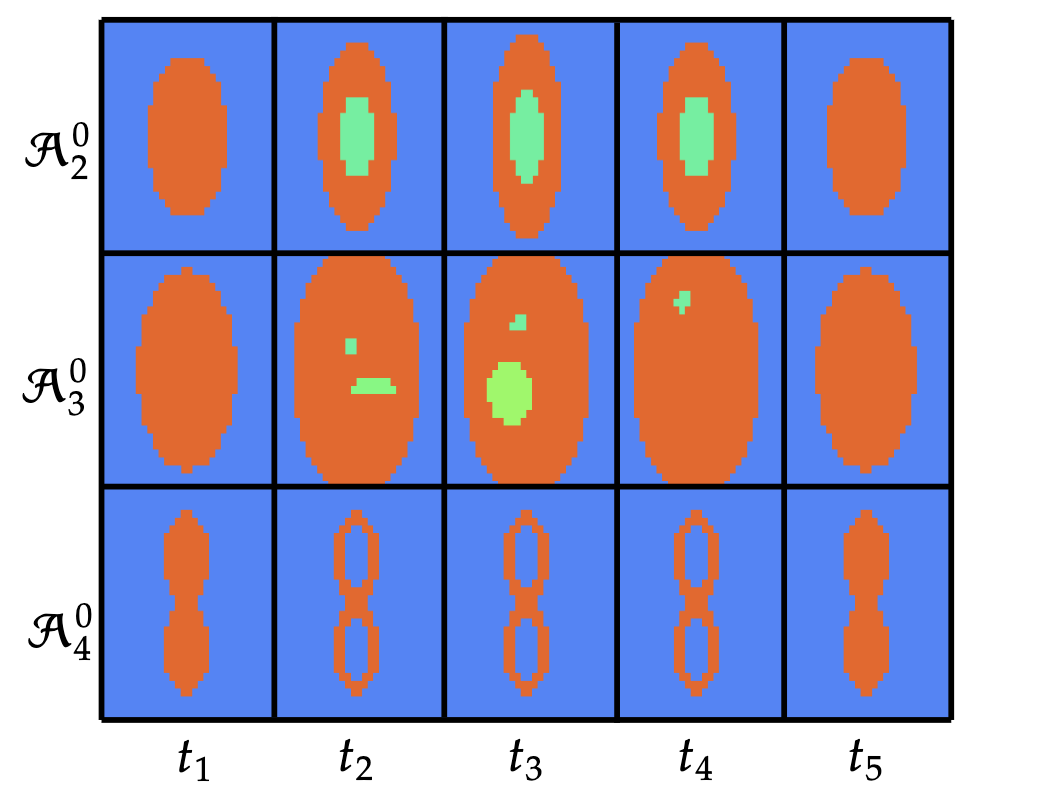}
 \caption{$\mathcal{A}_2^{0},\mathcal{A}_3^{0},\mathcal{A}_4^{0}$}\label{subfig:SimEx2}
 \end{subfigure}
 \begin{subfigure}{.05\linewidth}
 \centering
 \includegraphics[width=\linewidth]{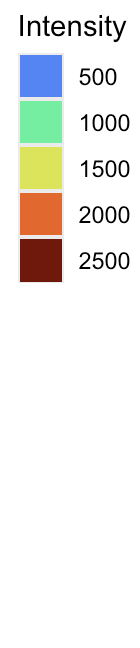}
 \end{subfigure}
 \begin{subfigure}{.3\linewidth}
 \centering
 \includegraphics[width=\linewidth]{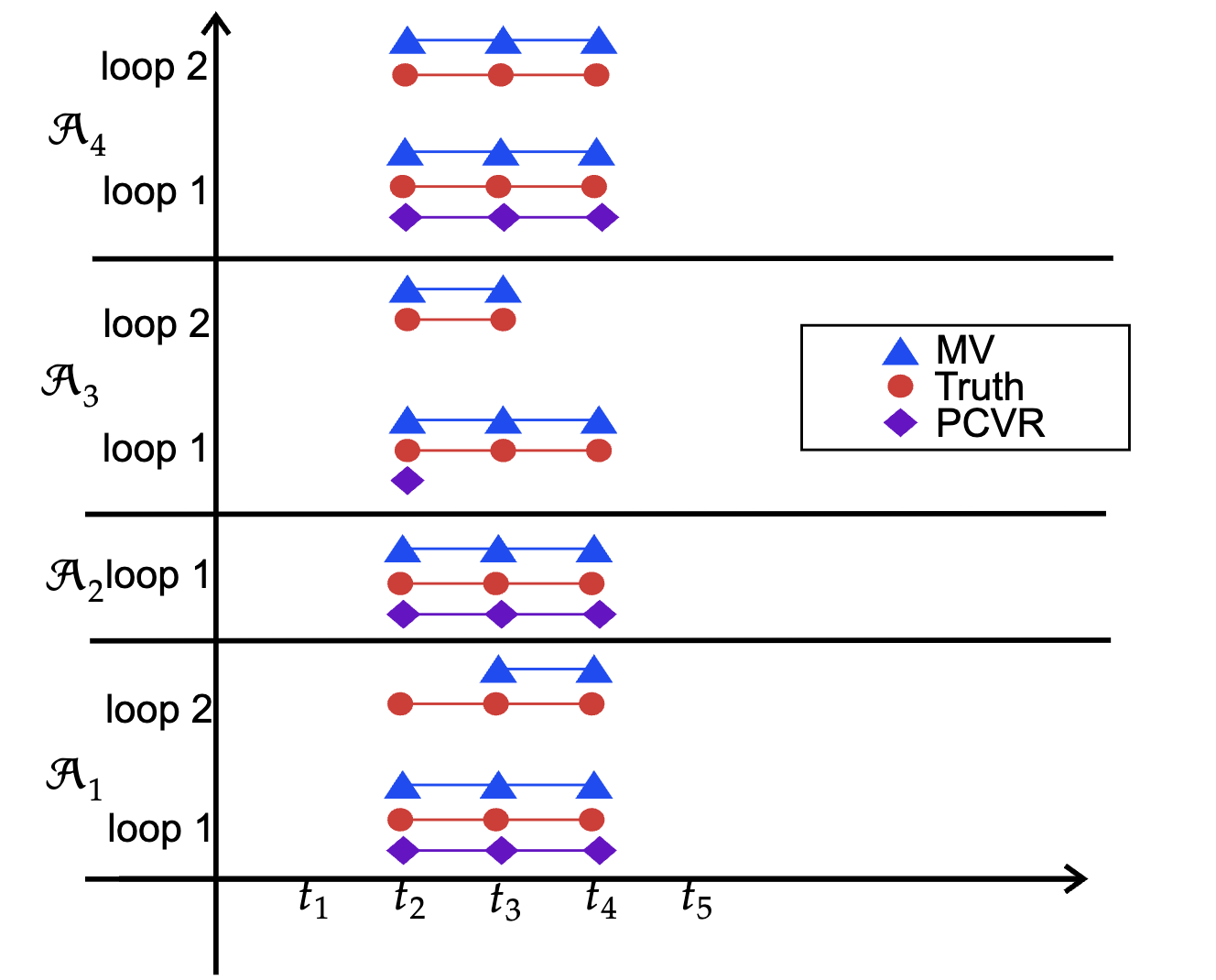}
 \caption{Results}\label{subfig:SimRes}
 \end{subfigure}
\caption{Simulation setup and results. (a) An example of an $\mathcal{A}_1$ time series where the first row is the true underlying pattern, the second row is a noisy realization of this underlying pattern, and the third row is a smoothed version of the noisy realization. (b) True loop pattern for the three other time series used in this study: $\mathcal{A}^{0}_2$ (first row), $\mathcal{A}^{0}_3$ (second row), $\mathcal{A}^{0}_4$ (third row). (c) The results where the $y-$axis lists all the true loops for each of the four examples. The points indicate if a loop was detected, and a line connecting points suggests the same loop was detected between time points. The truth is represented as red circles and lines, the proposed {\bf MV} method as blue triangles and lines, and the comparison {\bf PCVR} method as purple diamonds and lines.
} \label{fig:SimulationTimeSeries}
\end{figure}

The performance of the proposed {\bf MV} method was evaluated empirically and compared to the approach introduced in \cite{r2}, which we refer to as the {\em Point Cloud Vietoris–Rips} ({\bf PCVR}) method. The {\bf PCVR} framework has several approaches to evaluating topological features in time; we focus on the best performing approach for the numerical examples listed below. In this approach, each grayscale image in a time series is thresholded to create a binary image based on a predetermined value. These binary images are then converted to point clouds by placing a point at the center of any pixel with an intensity value of one, which, according to the {\bf PCVR} method, corresponds to a region of the binary image that contains a topological feature of interest. A Vietoris-Rips filtration is then applied to each point cloud to compute its persistent homology, where the simplicial complex construction is based on pairwise Euclidean distances between points, as opposed to the upper-level set filtration used in our method which is based on image intensity. At each point in time, a persistence diagram captures the topological features present which are then matched to features at neighboring time points based on persistence. The {\bf PCVR} method provides a useful benchmark for comparison, particularly because it has been previously applied to time series of cell wound images. However, it has limitations: it can lead to mismatches in feature correspondence over time and it only supports the tracking of an individual loop's persistences across time, similar to the approaches in \cite{r1} and \cite{r27}. In contrast, the {\bf MV} method is specifically designed to address these limitations. We highlight these situations below with the numerical examples seen in Figures~\ref{subfig:SimEx1} and \ref{subfig:SimEx2}. Performance is assessed through (i) the ability to identify loops at each time point that belong to the true pattern $\mathcal{A}^0$, and (ii) the ability to track these loops across time. 

Four distinct time series of images are used, denoted as $\mathcal{A}_1^{\sigma}, \mathcal{A}_2^{\sigma}, \mathcal{A}_3^{\sigma}, \text{ and } \mathcal{A}_4^{\sigma}$, where each one consists of five images with time points $t_o \in \{t_1, \ldots, t_5\}$. All of the time series contain at least one $H_1$ feature at some middle time point(s), as well as a connected component at both the initial and final time points. This setup is designed to simulate a cylindrical structure evolving through time, consistent with an $H_2$ feature (i.e., $\gamma_1$). Figure~\ref{subfig:SimEx1} shows an example of a time series of images $\mathcal{A}_1^{\sigma}$ generated from the true pattern $\mathcal{A}_1^{0}$, along with the corresponding smooth version of that times series $\mathcal{\tilde A}^{\sigma}_1$. Figure~\ref{subfig:SimEx2} shows the remaining three time series: $\mathcal{A}_2^{\sigma}, \mathcal{A}_3^{\sigma}, \text{ and } \mathcal{A}_4^{\sigma}$, which, in conjunction with $\mathcal{A}^{\sigma}_1$, are used to assess the ability of the {\bf MV} and {\bf PCVR} to represent the evolving loop structure over time. These data were designed to investigate each methods' performance in settings with multiple loops ($\mathcal{A}_1, \mathcal{A}_3, \mathcal{A}_4$), loops with thick boundaries ($\mathcal{A}_1,\mathcal{A}_2, \mathcal{A}_3$) versus thin boundaries ($\mathcal{A}_4$), or loops which are moving across time ($\mathcal{A}_3$).

The results of both method are shown in Figure~\ref{subfig:SimRes} where the $y-$axis lists the different loops within each time series ($\mathcal{A}_1$ loop 1, $\mathcal{A}_1$ loop 2, etc.) and the $x-$axis lists the different time points $t_1, \ldots, t_5$. A red dot is placed at each time point where a loop is present in the true underlying pattern, with red lines connecting consecutive time points if the same loop persists across them. The blue triangles and the connecting blue lines indicate instances where the {\bf MV} method correctly identified the presence and continuity of a loop over time. Similarly, the purple diamonds and the purple lines represent correct identifications and connections made by the {\bf PCVR} method. The goal is for the {\bf MV} and {\bf PCVR} methods to match the red dot and lines of the true underlying pattern.

In summary, the comparison {\bf PCVR} method accurately detected loops $50\%$ of the time while the proposed {\bf MV} method accurately detected loops in all but one instance (i.e., 95\% of the time). The primary reason for the improved performance of the {\bf MV} method is its use of an upper-level set filtration, which interprets persistence not in terms of the size of a hole, but as the difference in intensity values between the hole's boundary and its interior. In contract, the {\bf PCVR} method relies on the Vietoris–Rips filtration, which is based on pairwise distances. This makes it less effective at identifying small-area loops, which can be easily missed. A second advantage in the {\bf MV} method method is its use of a hypothesis testing framework to binarize the smoothed image data, rather than relying on a predefined threshold. This approach appears to produce a more accurate partitioning of loops. However, the {\bf PCVR} method was computationally faster than the {\bf MV} method, since it computes persistence diagrams over 2D data rather than the time-stacked 3D images for the {\bf MV} method. The {\bf MV} method performs well when an $H_2$ feature exists to connect lower-dimensional features across time. In particular, a cylindrical structure in the time series is necessary to determine an appropriate threshold for the 3D data (see Step 2 discussed in Section~\ref{subsec:step2}) so that the relevant $H_1$ features can be correctly connected. Although the persistence diagram computed from the 3D data in Step 2 does capture $H_1$ features, without the presence of an $H_2$ feature, these are not linked across time steps and some may be overlooked. The {\bf PCVR} method may be more comparable to the method created in \cite{r1} which can be used to track the persistence of an $H_1$ feature across time, though this method does not explicitly track features over time. The {\bf MV} method is especially effective at capturing meaningful changes to a loop structure across time outside of persistence, particularly when a single loop evolves into multiple loops, disappears, merges with other features, or changes dimension.

\section{Application to Wounded Cells}\label{sec:Data}

Patterns and pattern formation are prevalent and critically important features of biological systems. Patterns are evident across levels of biological organization, from the interior of cells to ecosystems, and they play essential roles in biology. During development, for example, factors that regulate gene expression accumulate in patterns that eventually subdivide the embryo into distinct segments \cite{c1}. During cell division, factors that control the machinery needed to split the cell in half accumulate in stripe-like patterns at the cell equator \cite{c2}. A particularly striking example of biological pattern formation is provided by the response of single cells to wounds \cite{c3}. Following damage to their surface, two proteins--Rho and Cdc42--are activated in concentric rings around the wounds, such that the ring of active Cdc42 circumscribes the ring of active Rho \cite{c4, c5}. These proteins, in turn, stimulate the assembly of actin filaments and myosin-2 around the wound edge \cite{c10}. The ring of actin filaments and myosin-2 contract, pulling the edges of the wound inward and thereby helping to close the wound \cite{c6, c7}. Thus, the patterns formed by Rho and Cdc42 are important for healing the cell, a point underscored by the observation that experimental perturbation of the Rho and Cdc42 rings perturbs the healing process \cite{c8, c9}. 
The proposed {\bf MV} method can be used to identify and quantify these rings patterns, and link the patterns across time.

\begin{figure}
\centering
\begin{subfigure}{.3\linewidth}
 \centering
 \includegraphics[width=\linewidth]{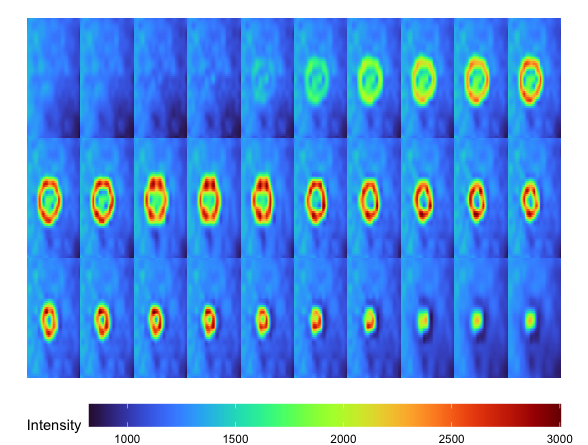}
 \caption{Video of $\mathcal{\tilde A}^{\text{C3}}_{t_o}$}\label{subfig:c3}
 \end{subfigure}
 \begin{subfigure}{.3\linewidth}
 \centering
 \includegraphics[width=\linewidth]{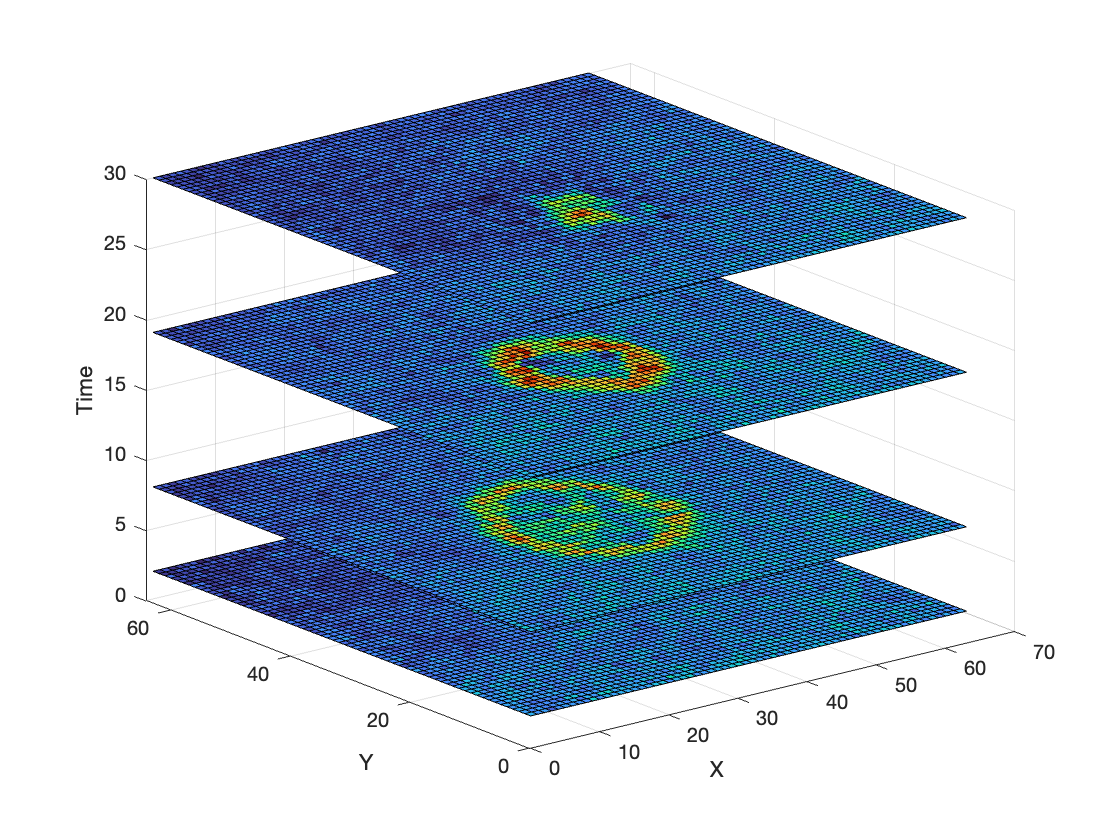}
 \caption{3D array of $\mathcal{\tilde A}^{\text{C3}}$}\label{subfig:c3Stack}
 \end{subfigure}
 \begin{subfigure}{.3\linewidth}
 \centering
 \includegraphics[width=\linewidth]{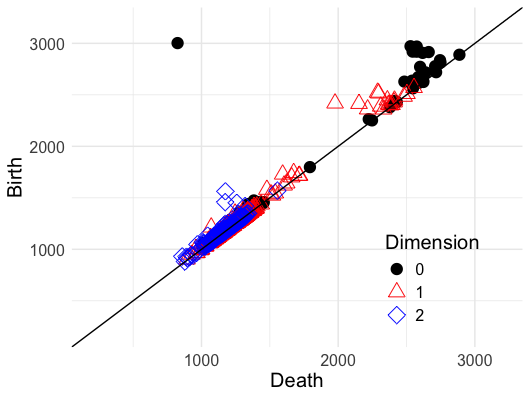}
 \caption{$\mathcal{P}(\mathcal{\tilde A}^{\text{C3}})$}\label{subfig:PDc3}
 \end{subfigure}
\caption{(a) Time series of images of the C3 cell where each row is a sequence of ten consecutive images (e.g. row one is $t_1-t_{10}$, row two is $t_{11}-t_{20}$, and row 3 is $t_{21}-t_{30}$). (b) The time series of images transformed into an array $\mathcal{\tilde A}^{\text{C3}}$ by adding a time dimension on the z-axis. (c) Persistence diagrams for $\mathcal{\tilde A}^{\text{C3}}$ where the color and shape of points are for different dimensional features $m=\{0,1,2\}$ and the death time is one the $x-$axis and the birth time is on the $y-$axis.} \label{fig:TimeSeries}
\end{figure}

The data for this analysis are a time series of smoothed images of different cells which have been wounded \cite{r1}. One is a Control cell which is only wounded and the other is a C3 cell which is wounded and injected with a toxin. A ring begins to appear around time $t_5$ and we want to track the evolution of the pattern in the wound for both the C3 and Control cells. In this analysis, the image resolution has been reduced to make computations less intensive and the images are smoothed to reduce noise and the bias in the birth/death times of the topological features as outlined in \cite{r1}. An example of the data can be seen in Figure~\ref{subfig:c3} which is the wound of the C3 cell over time.

In Step 1 of the {\bf MV} method a temporal dimension is added to the data and then $H_2$ features are identified in the stack of images for the C3 cell which can be seen in Figure~\ref{fig:TimeSeries}. Figure~\ref{subfig:c3} is the time series of smoothed images ($\{ \mathcal{\tilde A}^{\text{C3}}_{t_1}, \ldots, \mathcal{\tilde A}^{\text{C3}}_{t_{30}} \}$) which are transformed into the array $\mathcal{\tilde A}^{\text{C3}}$ in Figure~\ref{subfig:c3Stack} where only four slices in time of the array are shown for visualization purposes. The $H_2$ feature has a clear cylindrical form throughout the stack of images due to the cell wounds from the individual time points. The persistence diagram of the upper-level set filtration of $\mathcal{\tilde A}^{\text{C3}}$ is shown in Figure~\ref{subfig:PDc3} with $H_2$ features (blue diamonds), $H_1$ features (red triangles), and $H_0$ features (black dots). The birth time is on the $y-$axis and the death time is on the $x-$axis. The C3 cell has several persistent $H_2$ features on the persistence diagram (blue diamonds located further from the birth=death line), whereas the Control cell had only one prominently persistent $H_2$.

We assume that the $H_2$ feature ($\gamma_1$) and the partitions $\mathcal{G}(\boldsymbol{t^*})$ representing the wound in the arrays $\mathcal{\tilde A}^{\text{C3}}$ and $\mathcal{\tilde A}^{\text{Con}}$ follow the Assumptions~\ref{as:PersistenceWoundTime} and \ref{as:FunctionAcrossTime}. Therefore, the second step in the {\bf MV} method applies the Maximum Persistence Test on the entire array to identify the possible statistically significant $H_2$ feature, which represents the cell wound at potentially multiple points in time, instead of on each image individually. Steps 1-4 of Algorithm~\ref{alg:GetH2} outline how to apply the Maximum Persistence Test to the C3 and the Control cell arrays $\mathcal{\tilde A}^{\text{C3}}$ and $\mathcal{\tilde A}^{\text{Con}}$, respectively. Using $m=2$ for the dimension of the $H_m$ feature in the Maximum Persistence Test, each column of Table~\ref{tab:DataResults} shows the p-value$_{\text{max}}$, the 95th percentile of the null distribution $\boldsymbol{\rho^*}_{\textbf{max}}(2)$, and the persistence of the $H_2$ feature for both the C3 cell (row 1) and the Control cell (row 2). Since the most persistent $H_2$ features in $\mathcal{\tilde A}^{\text{C3}}$ and $\mathcal{\tilde A}^{\text{Con}}$ are statistically significant, the birth times of these features ($\hat \theta_{1}^{\text{C3}}$ and $\hat \theta_{1}^{\text{Con}}$) can be used to identify the part of the array where the wound is present ($\mathcal{G}_1^{\text{C3}}(\boldsymbol{t^*})$ and $\mathcal{G}_1^{\text{Con}}(\boldsymbol{t^*})$).
\begin{table}
\centering \small%
\begin{tabular}{|p{1.75cm}|p{1.75cm}|p{3.75cm}|p{2cm}|}
\hline
\textbf{Cell Type } & \textbf{p-value$_{\text{max}}$} & \textbf{$95^{\text{th}}$ percentile $\boldsymbol{\rho}^*_{\text{max}}(2)$} & \textbf{persistence} \\
\hline
\textbf{C3} & 0 & 99 & 388 \\
\hline
\textbf{Control} & 0 & 89 & 693\\
\hline
\end{tabular}
\captionof{table}{Results of the Maximum Persistence Test applied to the smoothed arrays where row is the cell type. The columns are the p-value$_{\text{max}}$ from the Maximum Persistence Test, the $95^{\text{th}}$ percentile of the null distribution of no structure in the array, and the persistence of the most persistent $H_2$ feature.} \label{tab:DataResults}
\end{table}

\begin{figure}
\centering
\begin{subfigure}{.45\linewidth}
 \centering
 \includegraphics[width=\linewidth]{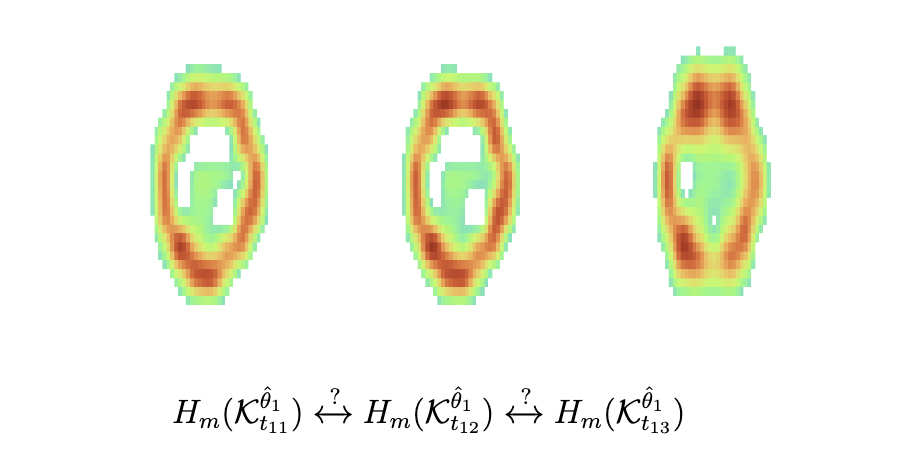}\caption{Slices of the cell wound}\label{subfig:slices_data}
 \end{subfigure}
 \begin{subfigure}{.45\linewidth}
 \centering
 \includegraphics[width=\linewidth]{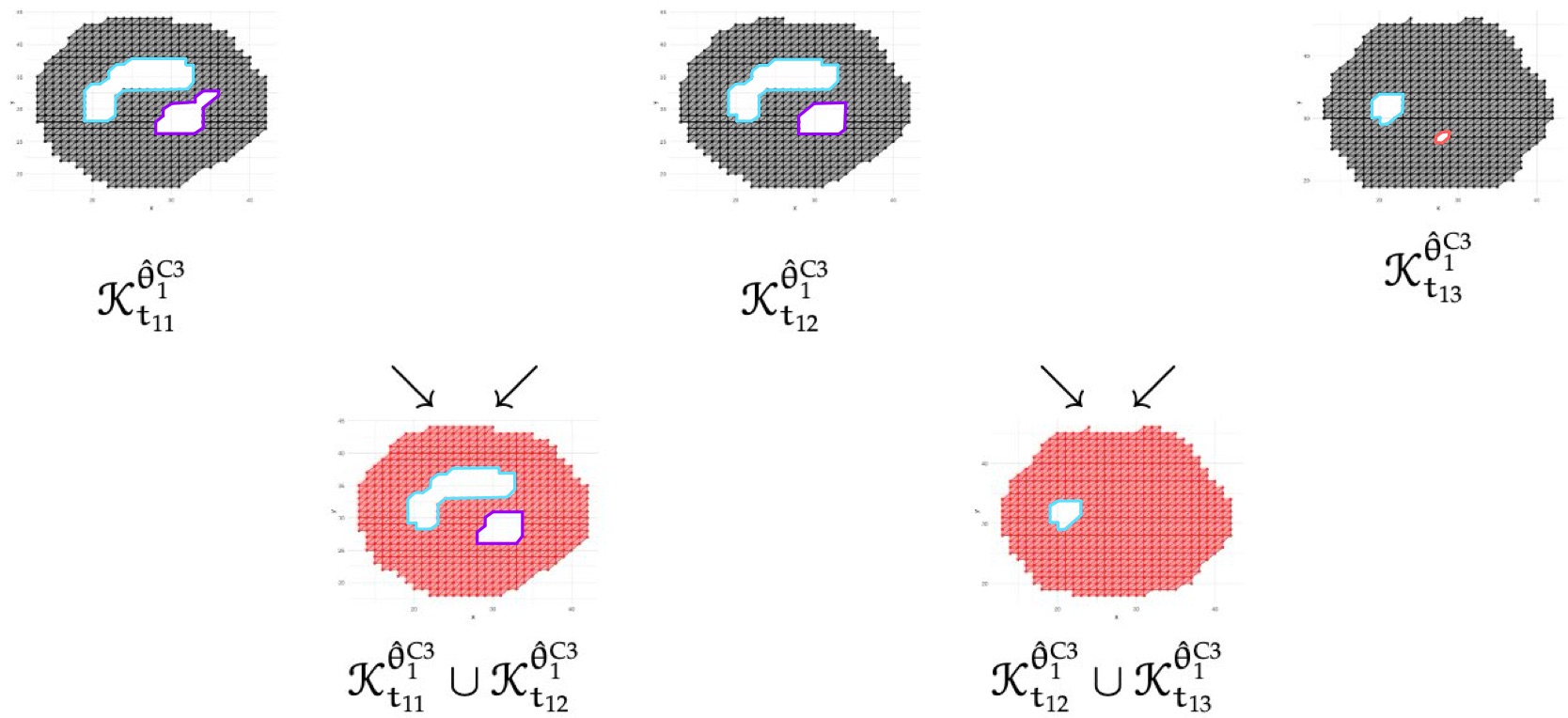}
 \caption{Alternating simplicial complexes}\label{subfig:sc_data}
 \end{subfigure}
\caption{(a) Slices of the cell wound $\mathcal{\hat G}_1^{\text{C3}}(t_{11})$, $\mathcal{\hat G}_1^{\text{C3}}(t_{12})$, $\mathcal{\hat G}_1^{\text{C3}}(t_{13})$ where color is the intensity value of the pixels which are above the threshold $\hat \theta_1^{\text{C3}}$. (b) The black simplicial complexes are the simplicial complexes representing the wound at time points $t_{11}, t_{12}, t_{13}$ and the red simplicial complexes are the unions.} \label{fig:ExampleZigZag_data}
\end{figure}

The third step of the {\bf MV} method identifies $\mathcal{\hat G}_1^{\text{C3}}(\boldsymbol{t^*_{\text{C3}}}))$ and $\mathcal{\hat G}_1^{\text{Con}}(\boldsymbol{t^*_{\text{Con}}}))$ from $\hat \theta_{1}^{\text{C3}}$ and $\hat \theta_{1}^{\text{Con}}$. Then the simplicial complexes at each point in time $t_o \in \boldsymbol{t^*_{\text{C3}}}$ or $t_o \in \boldsymbol{t^*_{\text{Con}}}$ ($\mathcal{K}^{\hat \theta_1^{\text{C3}}}_{t_o}$ and $\mathcal{K}^{\hat \theta_1^{\text{Con}}}_{t_o}$) are found from $\mathcal{\hat G}_1^{\text{C3}}(t_o)$ and $\mathcal{\hat G}_1^{\text{Con}}(t_o)$. Step 5 of Algorithm~\ref{alg:GetH2} describes how to get the list of simplicial complexes at each point in time $\mathcal{K}^{\hat \theta_1^{\text{C3}}}_{t_i}$ and $\mathcal{K}^{\hat \theta_1^{\text{Con}}}_{t_o}$. Examples of $\mathcal{\hat G}_1^{\text{C3}}(t_o)$ for times $t_{11}, t_{12}, t_{13}$ can be seen in Figure~\ref{subfig:slices_data} and the corresponding simplicial complexes $\mathcal{K}^{\hat \theta_1^{\text{C3}}}_{t_o}$ for times $t_{11}, t_{12}, t_{13}$ can be seen in Figure~\ref{subfig:sc_data} in black. The color of each loop indicates the grouping across time using zigzag persistence. The induced alternating structure of the persistence modules of $H_m(\mathcal{K}_{t_o}^{\hat \theta_1^{\text{C3}}})$ through unions are the red simplicial complexes.

\begin{figure}
\centering
 \begin{subfigure}{.45\linewidth}
 \centering
 \includegraphics[width=\linewidth]{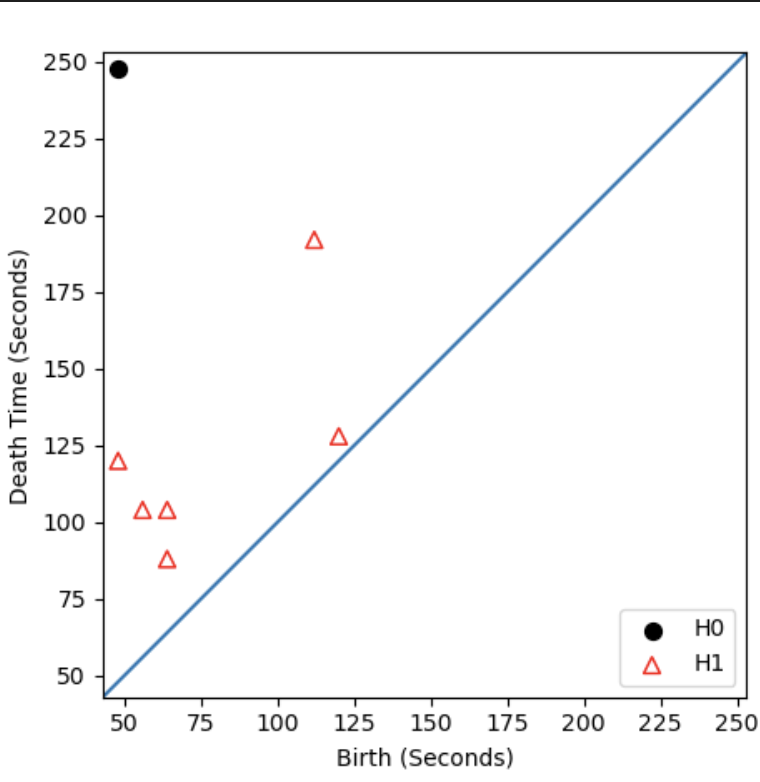}
 \caption{Zigzag $\mathcal{P}(\mathcal{\tilde A}^{\text{C3}})$}\label{subfig:ZZc3}
 \end{subfigure}
 \begin{subfigure}{.45\linewidth}
 \centering
 \includegraphics[width=\linewidth]{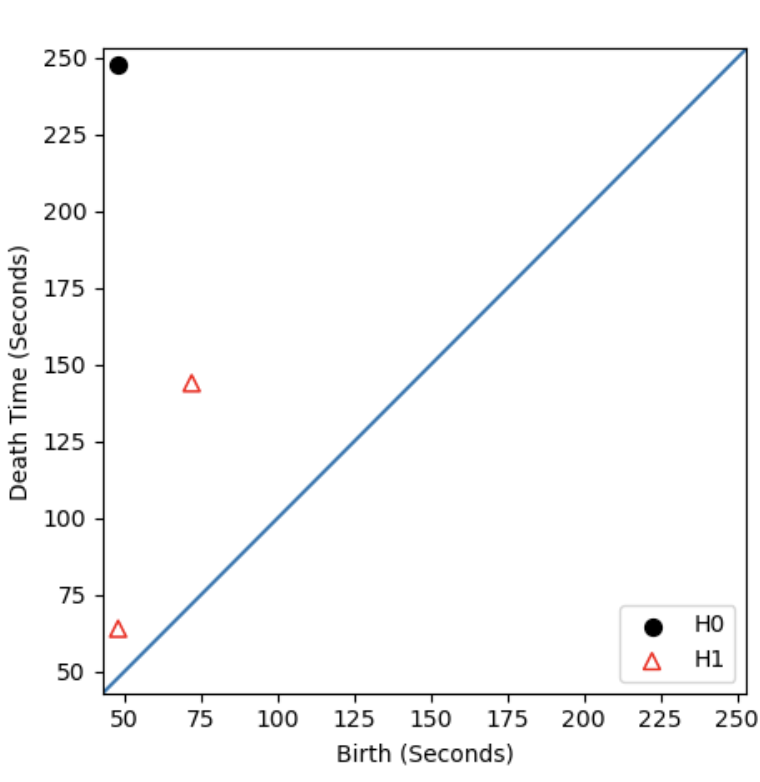}
 \caption{Zigzag $\mathcal{P}(\mathcal{\tilde A}^{\text{Con}})$}\label{subfig:ZZcon}
 \end{subfigure}
\caption{Zigzag diagrams for the array of the C3 cell (a) and the array of the Control cell. The red triangles are the $H_1$ features and the black dots are the $H_0$ features.}\label{fig:ZZtime}
\end{figure}

In Figure~\ref{fig:ZZtime}, the zigzag diagrams are shown where the birth ($x-$axis) and death ($y-$axis) times are displayed in seconds (e.g., $t_{12}=96$ seconds). The C3 cell is shown in Figure~\ref{subfig:ZZc3} and the Control cell is shown in Figure~\ref{subfig:ZZcon}. In general, the Control cell is more organized and consistent with only two loops which make up the wound. The C3 cell has much more disorganization in the wound with multiple loops and connected components making up the wound at different points in time. This method allows for a different way to track loops across time and to understand the patterns of complex time series.

\section{Discussion and Conclusion}

The proposed {\bf MV} method is designed to identify and track different dimensional holes in a time series of images using TDA. It has two primary differences compared to current methods that improve the soundness of the feature identification and tracking components. First, methods which apply TDA to images typically either convert the image to a graph, point cloud, or binary image, which requires a threshold value to be selected. The threshold is based on a arbitrary, predefined number to take the pixel intensity dimension out of the problem. In the proposed {\bf MV} method, we use a statistically-driven method for defining the threshold using a hypothesis testing framework. Second, the proposed {\bf MV} method includes time directly in the data structure instead of adding time to topological summary statistics (e.g., persistences, Betti numbers) in order to improve the tracking of the features across time. When tracking of features across time is done on the summary statistics such as persistence, errors can happen if the ordering of the features by persistence changes at any time point. Including time into the data structure facilitates the feature tracking across time.

For the cell wound application, the proposed {\bf MV} method captures the pattern in a sequence of images by tracking changes in the $H_1$ and $H_0$ features that constitute the wound across time. In \cite{r1}, a method that does not directly incorporate time found that the C3 cell exhibits higher persistence in the ring structure of the wound compared to the Control cell at earlier time points. However, this relationship changed at later time points, where the ring of the wound in the Control cell became more persistent. The {\bf MV} method shows that the C3 cell has more topological features representing the wound throughout the time series compared to the Control cell, as illustrated in Figure~\ref{subfig:ZZc3}. The {\bf MV} method provides additional information for quantifying cell wound patterns and approaches time series analysis.

The proposed {\bf MV} method has some limitations. First, calculating persistence diagrams for higher-dimensional features (e.g., $m=2$) is more computationally intensive than when time is excluded from the data structure. If smoothing of the images is needed, that also adds to the computational time. We mitigated this limitation in our work by reducing the resolution of the image; however, reducing the resolution may not work well in other applications so more computationally efficient methods are needed. One potential solution is using cubical homology which fits the underlying structure of an image better and is more computationally efficient compared to simplicial homology. Second, the assumptions on the data for the proposed method to be effective require a higher-dimensional topological feature connecting the lower-dimensional ones. In the cell wound application, the assumptions are satisfied since there is a clear cylindrical cavity formed in the stack of images. Other applications may, for example, have an $H_1$ feature appear at some point in time, but not close so that an $H_2$ feature does not form resulting in the proposed {\bf MV} method not detecting or tracking that $H_1$ feature. A possible generalization of this method to address this limitation could consider spatial information at consecutive time points to try and connect features across space and time instead of using a higher-dimensional feature. When trying to track the evolution of topological features over time, many open questions remain.




\begin{funding}
SG and JCK gratefully acknowledge support from NSF under Grant Number DMS 2038556. JCK gratefully acknowledges support from NSF under Grant Number 2337243. JZ gratefully acknowledges support from NSF under Grant Number DMS 2245906. WMB gratefully acknowledges support from NIH under Grant Number RO1 GM052932. Research presented in this article was supported by the National Security Education Center (NSEC) Informational Science and Technology Institute (ISTI) using the Laboratory Directed Research and Development program of Los Alamos National Laboratory under project number 20240479CR-IST
\end{funding}





\end{document}